%% file: ieee4double.tex
\documentclass[twocolumn,final]{IEEEtran}

\usepackage{graphicx}
\usepackage{amsmath}
\usepackage{amssymb}
\usepackage{cite}

\graphicspath{{figures/}}

\newcommand{\RR}{{\ensuremath{\mathbb R}}}
\newcommand{\ZZ}{{\ensuremath{\mathbb Z}}}
\newcommand{\NN}{{\ensuremath{\mathbb N}}}
\newcommand{\LL}{{\ensuremath{\mathrm L}}}
\newcommand{\HH}{{\ensuremath{\mathrm H}}}
\newcommand{\E}{\mathsf{E}}
\newtheorem{prop}{Proposition}

\begin{document}
\title{Image Analysis Using a Dual-Tree $M$-Band Wavelet Transform
\thanks{Part of this work was presented at the 2004 EUSIPCO conference \cite{Chaux_C_2004_p-eusipco_hilbert_pmbowb} and at the 2005 ICASSP conference \cite{Chaux_C_2005_p-icassp_2D_dtmbwd}.}}
\author{Caroline Chaux, {\em Student Member}, Laurent Duval, {\em Member} and\\
Jean-Christophe Pesquet, {\em Senior Member, IEEE}
\thanks{C. Chaux and J.-C. Pesquet are with the Institut Gaspard Monge and CNRS-UMR 8049,
Universit{\'e} de Marne la Vall{\'e}e, 77454 Marne la Vall{\'e}e Cedex 2, France.
E-mail: \texttt{\{chaux,pesquet\}@univ-mlv.fr}.}
\thanks{L. Duval is with the Institut Fran{\c c}ais du P{\'e}trole, 
Technology, Computer Science and Applied Mathematics  Department,
92500 Rueil Malmaison, France.
E-mail: \texttt{laurent.duval@ifp.fr}.}}

\maketitle
\vspace{-1.5 cm}
\begin{abstract}
We propose a 2D generalization to the $M$-band case of the dual-tree 
decomposition structure (initially proposed
by N. Kingsbury and further investigated by I. Selesnick) based on a Hilbert pair of wavelets. We particularly address (\textit{i}) the construction of the dual basis and (\textit{ii})  the resulting directional analysis. We also revisit the necessary pre-processing stage in the $M$-band case. While several reconstructions are possible because of the redundancy of the representation, we propose a new optimal signal reconstruction technique, which minimizes potential  estimation errors.
The effectiveness of the proposed $M$-band decomposition is demonstrated
via  denoising comparisons on several image types (natural, texture, seismics), with various $M$-band wavelets and thresholding strategies. Significant improvements in terms of both overall noise reduction and direction preservation are observed.
\end{abstract}


\section{Introduction}
The classical discrete wavelet transform (DWT) provides a means of implementing a multiscale analysis, based on a critically sampled filter bank with perfect reconstruction. It has been shown to be very effective both theoretically and practically \cite{Mallat_S_1998_book_wav_tsp} in the processing of  certain classes of signals, for instance piecewise smooth signals, having a finite number of discontinuities.  But, while  decimated transforms yield good compression performance, other data processing applications (analysis, denoising, detection) often require more sophisticated schemes than DWT.

One first drawback usually limiting the practical performance of DWT algorithms is their shift-variance with respect to the value of the transformed coefficients at a given scale. It often results in shift-variant edge artifacts at the vicinity of jumps, which are not desirable in real-world applications, signal delays being rarely known. 

A second  drawback  arises in dimensions greater than one: tensor products of standard wavelets usually possess poor directional properties. The later problem is  sensitive in feature detection or denoising applications. 
A vast  majority of the proposed solutions relies  on adding some redundancy to the transform. Redundancy based on shift-invariant  wavelet transforms
(see \cite{Nason_G_1995_incoll-was_sta_wtsa,Pesquet_J_1996_j-ieee-tsp_time-invariant_owr} and references therein) suppresses
  shift dependencies, at the expense of  an increased computational cost, which  often becomes intractable in higher dimensions.
 Less computationally-expensive approaches  have been developed on  complex filters for real signals
(we refer to \cite{Zhang_X_1999_j-ieee-tsp_orthogonal_cfbwpd} for an overview and
design examples), or by employing other wavelet frames \cite{Simoncelli_E_1992_tit_shi_mst}. For
instance, it is possible to resort to the concatenation of several
wavelet bases. One of the most promising decomposition
is  the \emph{dual-tree} discrete wavelet transform, proposed by
N. Kingsbury \cite{Kingsbury_N_2001_j-acha_complex_wsiafs}:
 two classical wavelet trees are developed in parallel, with 
 filters forming (approximate) Hilbert pairs.  Advantages of Hilbert
pairs had been earlier recognized by other authors \cite{Abry_P_1994_p-stfts_multiresolution_td}. In the complex case, the
 resulting analysis yields a redundancy of only $2^d$ for
 $d$-dimensional signals, with a much lower shift sensitivity and  better directionality in 2D  than the DWT.
The design of dual-tree filters is  addressed in 
\cite{Selesnick_I_2001_j-ieee-spl_hilbert_tpw}
through  an approximate Hilbert pair formulation 
for the ``dual'' wavelets. 
I.~Selesnick also proposed the double-density DWT and   combined both frame approaches 
\cite{Selesnick_I_2004_j-ieee-tsp_double-density_dtdwt}. The  \emph{phaselet} extension of the
dual-tree DWT has  been recently
introduced by R. Gopinath in
\cite{Gopinath_R_2003_tsp_pha_tirnsiwt}.
More recently, several authors have also proposed a projection scheme
with  an explicit control of the redundancy or with specific filter bank structures
\cite{Fernandes_F_2004_icassp_non_rlpsodcw,VanSpaendonck_R_2003_p-icassp_ort_htfbw}. 
Finally, other works on  the blending of analytic signals  and wavelets must be mentioned
\cite{Olhede_S_2003_tr_ana_wt,Chan_W_2004_icassp_dir_hwmsap}, in the context
of denoising or higher dimension signal processing. Recent developments based on ``geometrical'' wavelets are not mentioned here, in spite of their relevance.

A third drawback concerns design limitations in two-band decompositions: orthogonality, realness, symmetry, compactness of the  support and other properties (regularity, vanishing moments) compete.  The relative sparsity of good filter banks amongst all possible solutions is also well-known. In order to improve 
both design freedom and filter behavior, $M$-band filter banks and wavelets have been  proposed  \cite{Malvar_H_1992_book_signal_plt,Steffen_P_1993_j-ieee-tsp_theory_rmbwb,Tran_T_2000_tsp_lin_pprfblsdaic}. 

Improving on our previous work \cite{Chaux_C_2004_p-eusipco_hilbert_pmbowb},  we propose the construction of a 2D dual-tree $M$-band wavelet decomposition.   The organization of the paper is as follow:
in Section \ref{sec:hilbert_pair}, we investigate the theoretical conditions for the construction of   $M$-band Hilbert pairs.
In Section \ref{sec:2Dtransform}, we extend previous results on the pre-processing stage to the $M$-band context and illustrate  the direction extraction with the constructed wavelets. Since several reconstructions are possible, due to the decomposition redundancy, we then propose an optimal pseudo-inverse based frame reconstruction, which allows to reduce the effects of  coefficient estimation errors. Implementation issues are discussed
in Section \ref{se:implement}. In  Section \ref{sec:denoise},
we consider image denoising applications and
provide experimental results
showing significant improvements in terms of both noise reduction and direction preservation. 
Conclusions are drawn in Section \ref{sec:conclu}.

\section{Construction of $M$-band Hilbert pairs}
\label{sec:hilbert_pair}
\subsection{Problem statement}
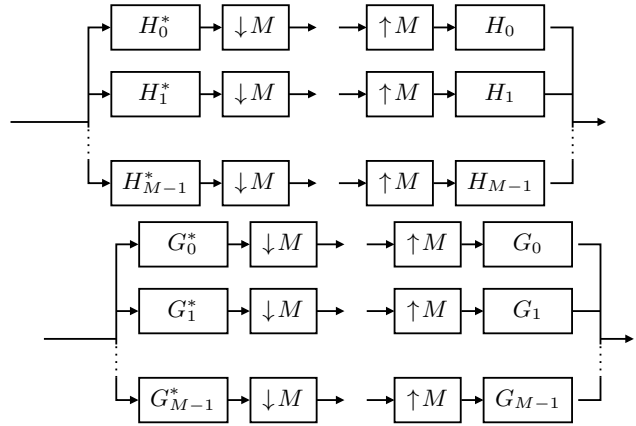
\begin{figure}
\centering
\input{figures/fb_M_ortho_dual.pstex_t}

\caption{A pair of analysis/synthesis $M$-band para-unitary filter banks.
}
\label{fig:Mband}
\end{figure}

In this section, we will focus on 1D signals belonging to the space $\LL^2(\RR)$
of square integrable functions.
Let $M$ be an integer greater than or equal to 2.
Recall that an $M$-band multiresolution analysis of $\LL^2(\RR)$
is defined by
one scaling function (or father wavelet) $\psi_0 \in \LL^2(\RR)$ and
$(M-1)$ mother wavelets $\psi_m \in  \LL^2(\RR)$, $m\in \{1,\ldots,M-1\}$
\cite{Steffen_P_1993_j-ieee-tsp_theory_rmbwb}.
These functions are solutions of the following scaling equations:
\begin{multline}
\forall m \in \{0,\ldots,M-1\},\\
\frac{1}{\sqrt{M}}\psi_m(\frac{t}{M})
= \sum_{k=-\infty}^\infty h_m[k] \psi_0(t-k),
\label{eq:twoscale}
\end{multline}
where the sequences $(h_m[k])_{k\in \ZZ}$ are
square integrable. In the following, we will assume
that these functions (and thus the associated
sequences $(h_m[k])_{k\in \ZZ}$)) are real-valued.
The Fourier transform of $(h_m[k])_{k\in \ZZ}$
is a $2\pi$-periodic function, denoted by
$H_m$. Thus, in the frequency domain, 
Eq. \eqref{eq:twoscale} can be re-expressed as:
\begin{equation}
\forall m \in \{0,\ldots,M-1\},\;
\sqrt{M}\widehat{\psi}_m(M\omega)
= H_m(\omega) \widehat{\psi}_0(\omega),
\label{eq:twoscalef}
\end{equation}
where $\widehat{a}$ denotes the Fourier transform of a function
$a$.
For the set of functions
$\cup_{m=1}^{M-1} \{M^{-j/2} \psi_m(M^{-j}t-k), (j,k) \in \ZZ^2\}$
to correspond to an orthonormal basis of $\LL^2(\RR)$,
the following para-unitarity conditions must hold:
\begin{multline}
\forall (m,m^\prime) \in \{0,\ldots,M-1\}^2,\\
\sum_{p=0}^{M-1} H_m(\omega+p \frac{2\pi}{M})
H_{m^\prime}^*(\omega+p \frac{2\pi}{M}) = M \delta_{m-m'},
\label{eq:paraunitarity}
\end{multline}
where $\delta_m = 1$ if $m=0$ and 0 otherwise. The filter with
frequency response $H_0$ is low-pass whereas usually the filter with frequency
response $H_m$, $m \in \{1,\ldots,M-2\}$ (resp. $m=M-1$) is band-pass
(resp. high-pass).
In this case,
cascading  the $M$-band para-unitary analysis and  synthesis
filter banks, depicted in the upper branch in Fig. \ref{fig:Mband}, allows us to decompose
and to  reconstruct perfectly a given signal.

Our objective is to construct a ``dual'' $M$-band multiresolution
analysis defined by a scaling function $\psi_0^\HH$
and mother wavelets $\psi_m^\HH$, $m\in \{1,\ldots,M-1\}$.
More precisely, the mother wavelets will be obtained
by a Hilbert transform from
the ``original'' wavelets $\psi_m$, $m\in \{1,\ldots,M-1\}$.
In the Fourier domain, the desired property reads:
\begin{equation}
\forall m \in \{1,\ldots,M-1\},\qquad
\widehat{\psi}_m^\HH(\omega) = - \imath\;\mathrm{sign}(\omega)
\widehat{\psi_m}(\omega),
\label{eq:Hilbertcond}
\end{equation}
where $\mathrm{sign}$ is the signum function defined as:
\begin{equation}
\mathrm{sign}(\omega) = \begin{cases}
1 & \mbox{if $\omega > 0$}\\
0 & \mbox{if $\omega = 0$}\\
-1 & \mbox{if $\omega < 0$.}
\end{cases}
\end{equation}
As it is common in wavelet theory, Eq. \eqref{eq:Hilbertcond}, as well as all equalities in the paper
involving square integrable functions, holds almost everywhere (that is,
for all $\omega \not\in \Omega$ where $\Omega$ is a real set of zero measure).

Furthermore, the functions $\psi_m^\HH$ are defined by
scaling equations similar to \eqref{eq:twoscale} involving real-valued sequences
$(g_m[k])_{k\in \ZZ}$:
\begin{align}
\forall m \in \{0,\ldots,M-1\},& \nonumber \\
\frac{1}{\sqrt{M}}\psi_m^\HH(\frac{t}{M})
&= \sum_{k=-\infty}^\infty g_m[k] \psi_0^\HH(t-k)
\label{eq:twoscaled}\\
\Longleftrightarrow  \sqrt{M} \widehat{\psi}_m^\HH(M\omega)
&= G_m(\omega) \widehat{\psi}_0^\HH(\omega).
\label{eq:twoscaledf}
\end{align}
 In order to generate a dual $M$-band orthonormal
wavelet basis of $\LL^2(\RR)$,
the Fourier transforms $G_m$ of the sequences $(g_m[k])_{k\in \ZZ}$ must
also satisfy the para-unitarity conditions:
\begin{multline}
\forall (m,m^\prime) \in \{0,\ldots,M-1\}^2,\\
\sum_{p=0}^{M-1} G_m(\omega+p \frac{2\pi}{M})
G_{m^\prime}^*(\omega+p \frac{2\pi}{M}) = M \delta_{m-m'}.
\label{eq:paraunitarityd}
\end{multline}
The corresponding para-unitary Hilbert filter banks are illustrated by the lower branch in Fig.~\ref{fig:Mband}.

\subsection{Sufficient conditions for obtaining dual decompositions}
The Hilbert condition  \eqref{eq:Hilbertcond} yields
\begin{equation}
\forall m \in \{1,\ldots,M-1\},\qquad 
|\widehat{\psi}_m^\HH(\omega)| = |\widehat{\psi}_m(\omega)|.
\end{equation}
If we further impose that $|\widehat{\psi}_0^\HH(\omega)| =
|\widehat{\psi}_0(\omega)|,$
the scaling equations \eqref{eq:twoscalef} and \eqref{eq:twoscaledf} lead to
\begin{equation}
\forall m\in \{0,\ldots,M-1\},\qquad
G_m(\omega) = e^{-\imath \theta_m(\omega)} H_m(\omega),
\label{eq:allpass}
\end{equation}
where $\theta_m$ is $2\pi$-periodic. The phase functions $\theta_m$ should also be odd (for real filters) and thus only need to be determined over $[0,\pi]$.

For any $(m,m^\prime) \in \{0,\ldots,M-1\}^2$ with $m<m^\prime$, let  $(\text{\textsf{P}}_{m,m'})$ denote the following assumption:
The function $\alpha_{m,m'} = \theta_{m^\prime}-\theta_m$
is such that, for (almost) all $\omega \in [0,2\pi[$,
\begin{equation}
\alpha_{m,m'}(\omega+\frac{2\pi}{M})
= \alpha_{m,m'}(\omega)\pmod{2\pi}.
\end{equation}

Assuming that  Eq. \eqref{eq:paraunitarity} is satisfied,
it is then straightforward to verify that
the para-unitarity conditions \eqref{eq:paraunitarityd} for the dual filter bank hold if $(\text{\textsf{P}}_{m,m'})$ holds.
We are then able to state the following result:
\begin{prop}
Assume that
Conditions \eqref{eq:allpass} hold.
A necessary and sufficient condition for Eq. \eqref{eq:Hilbertcond}
to be satisfied is that there exists $\tilde{\theta}_0 = \theta_0 \pmod{2\pi}$
such that
\begin{equation}
\beta(\omega) = \sum_{i=1}^\infty \tilde{\theta}_0\left(\frac{\omega}{M^i}\right)
\label{eq:defbeta}
\end{equation}
is a convergent series and, $\forall m \in \{1,\ldots,M-1\}$,
\begin{equation}
\tilde{\alpha}_{0,m}\left(\frac{\omega}{M}\right) + \beta(\omega)
= \frac{\pi}{2}
\mathrm{sign}(\omega) \pmod{2\pi}\label{eq:SelesM}
\end{equation}
where $\tilde{\alpha}_{0,m} = \theta_m - \tilde{\theta}_0$.
\end{prop}
\begin{proof}
Given that $\widehat{\psi}_0(0)=1$, for $m=0$
Eq. \eqref{eq:twoscalef} is equivalent to
\begin{equation}
\widehat{\psi}_0(\omega)=
\prod_{i=1}^\infty[\frac{1}{\sqrt{M}}H_0(\frac{\omega}{M^i})].
\label{eq:iterscal}
\end{equation}
Similarly, we have for the ``dual'' scaling function:
\begin{equation}
\widehat{\psi}_0^\HH(\omega)= \prod_{i=1}^\infty
[\frac{1}{\sqrt{M}}G_0(\frac{\omega}{M^i})].
\label{eq:iterscalH}
\end{equation}
Furthermore, the expressions of the Fourier transforms of the mother wavelets
and ``dual'' mother wavelets can be deduced from Eqs. \eqref{eq:twoscalef}
and \eqref{eq:twoscaledf}. Consequently, Condition \eqref{eq:Hilbertcond}
may be rewritten as $\forall m \in \{1,\ldots,M-1\}$,
\begin{multline}
G_m(\frac{\omega}{M}) \prod_{i=2}^\infty
[\frac{1}{\sqrt{M}}G_0(\frac{\omega}{M^i})]
= \\ - \imath\;\mathrm{sign}(\omega) H_m(\frac{\omega}{M})\prod_{i=2}^\infty[\frac{1}{\sqrt{M}}H_0(\frac{\omega}{M^i})].
\end{multline}
Using Eq. \eqref{eq:allpass}, we see that the above relation is verified if and only if there exists $\tilde{\theta}_0 = \theta_0 \pmod{2\pi}$ such that
\begin{multline*}
\forall m \in \{1,\ldots,M-1\},\\
\theta_m(\frac{\omega}{M})+\sum_{i=2}^\infty\tilde{\theta}_0(\frac{\omega}{M^i})=\frac{\pi}{2}\mathrm{sign}(\omega) \pmod{2\pi}
\end{multline*}
where the involved series is convergent.
The above equation is obviously equivalent to Eq. \eqref{eq:SelesM}.
\end{proof}

Eqs. \eqref{eq:SelesM} and \eqref{eq:defbeta} constitute a generalization to the $M$-band case
of a famous result by Selesnick \cite{Selesnick_I_2001_j-ieee-spl_hilbert_tpw} restricted to dyadic wavelets.
One can remark that the convergence properties of the series $\beta(\omega)$ are only related
to the behaviour of $\tilde{\theta}_0$ around the origin since
$\omega/M^i \to 0$ as $i\to \infty$.
It is also worth noting that the function $\beta$ is given by the following ``additive''
scaling equation:
\begin{equation}
\beta(\omega) = \beta\left(\frac{\omega}{M}\right)+
\tilde{\theta}_0\left(\frac{\omega}{M}\right).
\label{eq:addscale}
\end{equation}

\subsection{Linear phase solution} \label{se:linphase}
In the 2-band case (under weak assumptions), $\tilde{\theta}_0$ verifying
Eqs. \eqref{eq:SelesM} and \eqref{eq:defbeta} is a linear function on $[-\pi,\pi[$ \cite{Selesnick_I_2001_j-ieee-spl_hilbert_tpw}.
In the $M$-band case, we will slightly restrict this constraint on
a smaller interval by imposing:
\begin{equation}
\forall \omega \in [0,2\pi/M[,\qquad
\tilde{\theta}_0(\omega) = \gamma \omega,
\label{eq:linear}
\end{equation}
where $\gamma \in \RR$. This choice clearly guarantees that the series $\beta(\omega)$ is convergent.
Using Eq. \eqref{eq:addscale}, after some calculations which are provided in Appendix \ref{ap:phase}, the following result can be proved:
\begin{prop}
Under the three conditions \eqref{eq:allpass}, $(\text{\textsf{P}}_{0,m})_{m\geq 1}$ and \eqref{eq:linear},
the solutions (modulo $2\pi$) to Eq.~\eqref{eq:SelesM}  are  given by
\begin{multline}
\forall m \in \{1,\ldots,M-1\},\\
\tilde{\alpha}_{0,m}(\omega) = \begin{cases}
\displaystyle \frac{\pi}{2}-(d+\frac{1}{2})M \omega & \mbox{if $\displaystyle \omega \in ]0,\frac{2\pi}{M}[$,}\\
0 & \mbox{if $\omega = 0$.}
\label{eq:alpha0mlin}
\end{cases}
\end{multline}
and $\; \forall p \in \left\{0,\ldots,\Big\lceil  \frac{M}{2}\Big\rceil
  -1\right\}, \forall \omega \in \Big[p \frac{2\pi}{M},(p+1)\frac{2\pi}{M}\Big[$,
\begin{equation}
\tilde{\theta_0}(\omega) = (d+\frac {1}{2})(M-1)\,\omega - p\pi,
\label{eq:theta0}
\end{equation}
where $d \in \ZZ$ and $\lceil u \rceil$ denotes the upper integer
part of a real $u$.
\label{p:phase}
\end{prop}

\begin{figure} 
\centering
\includegraphics{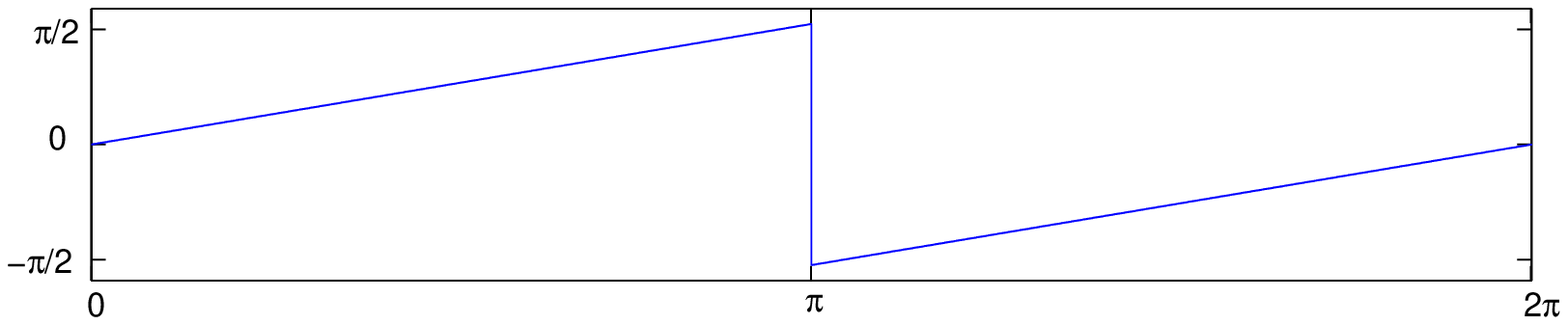}
\includegraphics{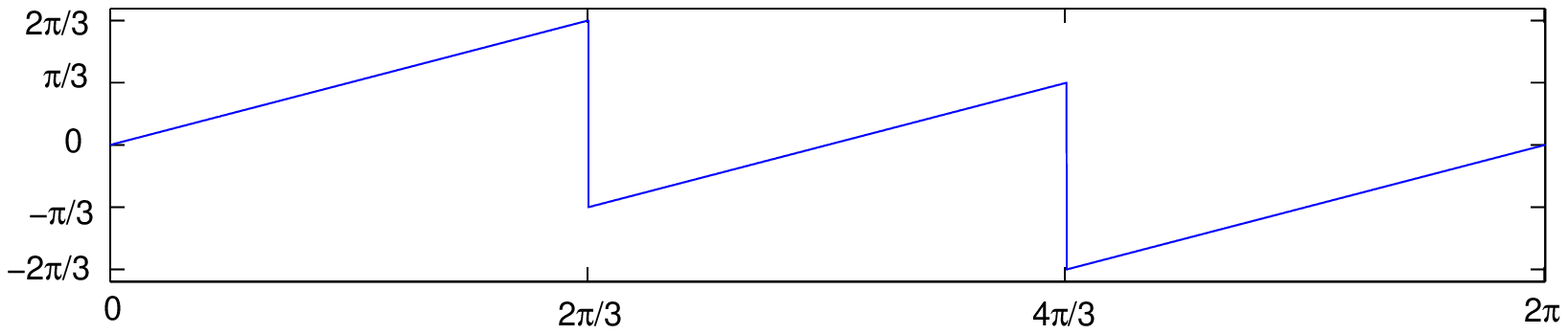}
\includegraphics{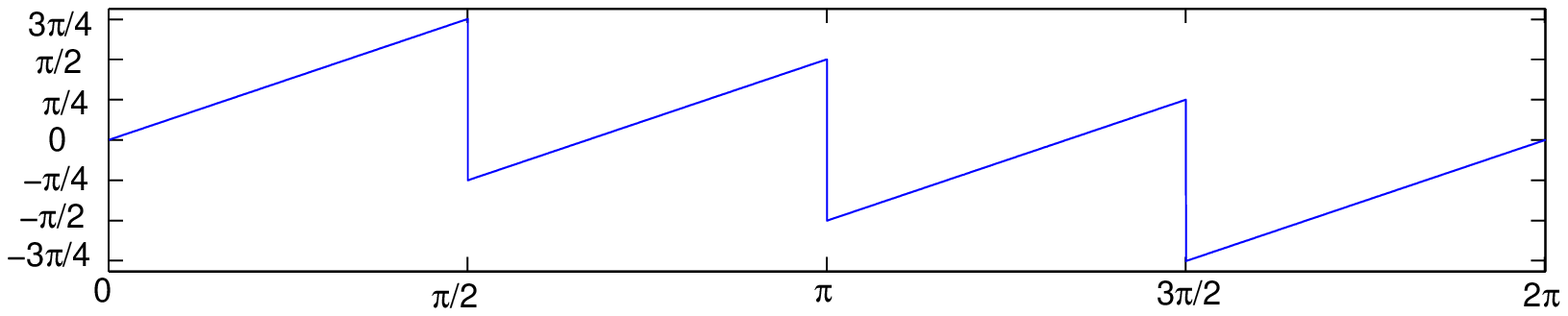}
\caption{Variations of $\tilde{\theta}_0(\omega)$ w.r.t. $\omega$,
for different numbers of channels: $M = 2$ (top), $M=3$ (middle) and $M=4$ (bottom).}
\label{fig:theta0M2}
\end{figure}

The integer $d$ defines a possible arbirtary delay between the filters of
the original and dual decompositions. Up to this delay,
Proposition   \ref{p:phase} states that, subject to \eqref{eq:allpass}, 
$(\text{\textsf{P}}_{0,m})_{m\geq 1}$ and \eqref{eq:linear}, there exists
a unique  solution to Eq.~\eqref{eq:SelesM}. It should also be noted that except for the 2-band case, $\tilde{\theta}_0$ exhibits discontinuities on $]0,\pi[$
 due to the $p\pi$ term (see Fig.~\ref{fig:theta0M2}).
These discontinuities however occur at zeros of the frequency response of the lowpass filter since we have
$H_0(2p\pi/M) = 0$, for all $p\in \{1,\ldots,M-1\}$
\cite{Steffen_P_1993_j-ieee-tsp_theory_rmbwb}.

We subsequently deduce the following corollary of the above proposition:
\begin{prop}
Para-unitary $M$-band Hilbert filter banks are obtained by choosing 
the phase functions defined by Eq.~\eqref{eq:theta0} and
\begin{multline}
\forall m\in \{1,\ldots,M-1\},\\ \theta_m(\omega) =
\left\{
\begin{array}{ll}
\displaystyle \frac{\pi}{2}-\left(d+\frac{1}{2}\right)\omega & \mbox{if $\omega \in ]0,2\pi[$},\\
0 & \mbox{if $\omega = 0$},
\end{array}
\right.
\label{eq:thetam}
\end{multline}
where $d \in \ZZ$. Then, the scaling function associated to the dual
wavelet decomposition is such that
\begin{multline}
\forall k \in \NN ,\;\forall \omega \in [2k\pi,2(k+1)\pi[,\\
\widehat{\psi}_0^\HH(\omega) =  (-1)^k e^{-\imath(d+\frac{1}{2})\omega}\;\widehat{\psi}_0(\omega).
\label{eq:linkpsi0Hpsi0}
\end{multline}
\end{prop}
\begin{proof}
It is readily shown that,
if $\tilde{\theta}_0$ is given by Eq.~\eqref{eq:theta0},
$\tilde{\alpha}_{0,m}$ is a $2\pi/M$-periodic function satisfying (almost everywhere) Eq.~\eqref{eq:alpha0mlin} if and only if  the functions $\theta_m$, $m \in \{1,\ldots,M-1\}$,
are expressed by Eq.~\eqref{eq:thetam} (modulo $2\pi$). Then,
we conclude from Proposition~\ref{p:phase} that the phases
given by Eqs.~\eqref{eq:theta0}-\eqref{eq:thetam} allow us to satisfy
the Hilbert condition \eqref{eq:SelesM}.
Furthermore, the functions  $\theta_m$, $m \in \{1,\ldots,M-1\}$, being all equal,
the paraunitary conditions $(\text{\textsf{P}}_{m,m'})_{m'>m\geq 0}$
are obviously fulfilled.
According to Eqs.~\eqref{eq:defbeta}, \eqref{eq:iterscal} and \eqref{eq:iterscalH}, $\widehat{\psi}_0^\HH(\omega) =  e^{-\imath\beta(\omega)}\;\widehat{\psi}_0(\omega)$. When
$\tilde{\theta}_0$ takes the form \eqref{eq:theta0},
the expression of $\beta$ is given by Eq.~\eqref{eq:inductbeta} in Appendix \ref{ap:phase}, thus yielding Eq.~\eqref{eq:linkpsi0Hpsi0}.
\end{proof}
Note that in the dyadic case, necessary and  sufficient conditions have been found for the linear phase property \cite{Ozkaramanli_H_2003_tsp_pha_cshtpwb}.

\subsection{Compact support}\label{se:compactsup}
Compactly supported wavelets are obtained with FIR (Finite Impulse
Response) filters. However, if the filters with frequency responses
$H_m(\omega)$ with $m\in \{1,\ldots,M-1\}$ are FIR (i.e.
$H_m(\omega)$ is a Laurent polynomial 
in $e^{\imath \omega}$),
the dual filters with frequency responses $G_m(\omega)$ cannot be FIR.
Indeed, the $\omega/2$ term in Eq.~\eqref{eq:thetam} prevents $G_m(\omega)$
from being a polynomial or even a rational function in $e^{\imath \omega}$. When $M$ is even, a similar argument holds showing that the low-pass filter $G_0(\omega)$ cannot be
FIR if  the primal one is FIR and Eq.~\eqref{eq:theta0} is satisfied.
When $M$ is odd, the jumps of $\pi$ arising for $\tilde{\theta}_0$
at frequencies $2p\pi/M$ with $p\in\{1,\ldots,M-1\}$ allow us to draw
the same conclusion. In other words, starting from orthonormal compactly
supported scaling functions/wavelets, it is not possible to generate
dual basis functions having a compact support.
However, the study of approximate FIR Hilbert pairs satisfying perfect reconstruction has been addressed by several authors in the dyadic case \cite{Tay_D_2004_icassp_sol_omcbwcp}, \cite{Kingsbury_N_2001_j-acha_complex_wsiafs}.

\subsection{Symmetry properties}
As already pointed out, one of the main advantage of the $M$-band
case with $M>2$ is to allow the construction of non-trivial real
orthonormal bases with compact support \textit{and} symmetric (or antisymmetric) wavelets.
Assume that symmetry properties are fulfilled for the primal filter bank. 
We now show that the dual filters and wavelets inherit these properties. Indeed, it can be proved (see Appendix \ref{ap:sym}) that:
\begin{prop} \label{prop:sym}
Let phase conditions \eqref{eq:theta0}, \eqref{eq:thetam} be satisfied.
If the low-pass impulse response
$(h_0[k])_{k\in \ZZ}$ is symmetric w.r.t. $k_0 \in \ZZ$, 
and, for $m \in \{1,\ldots,M-1\}$, $(h_m[k])_{k\in \ZZ}$ is symmetric (resp. antisymmetric)
w.r.t. $k_m \in \ZZ$, then
 $(g_0[k])_{k\in \ZZ}$ is symmetric w.r.t. $k_0+(d+\frac{1}{2})(M-1)$
and $(g_m[k])_{k\in \ZZ}$ is antisymmetric (resp. symmetric)
w.r.t. $k_m-d-\frac{1}{2}$.
\end{prop}
Under the assumption of the above proposition, Eqs.~\eqref{eq:iterscal}
and \eqref{eq:twoscalef}
allow us to claim that $\psi_0$ is symmetric w.r.t.
\begin{equation}
\tau = \frac{k_0}{M-1}
\end{equation}
and, for $m\in \{1,\ldots,M-1\}$, $\psi_m$ is
symmetric (resp. antisymmetric) w.r.t. $(\tau+k_m)/M$. Then, it is easily deduced from Eqs.~\eqref{eq:linkpsi0Hpsi0} and \eqref{eq:Hilbertcond} that $\psi_0^{\mathrm{H}}$
is symmetric w.r.t. $\tau+d+1/2$ and, for $m\in \{1,\ldots,M-1\}$,
$\psi_m^{\mathrm{H}}$ is  antisymmetric (resp. symmetric) w.r.t. $(\tau+k_m)/M$.

\section{Extension to 2D dual-tree $M$-band wavelet analysis}
\label{sec:2Dtransform}

\subsection{2D Decomposition}
\label{se:generality2D}

Two-dimensional separable $M$-band wavelet bases can be deduced from
the 1D dual-tree decomposition derived in Section \ref{sec:hilbert_pair}.
The so-obtained bases of $\LL^2(\RR^2)$ (the space
of square integrable functions defined on $\RR^2$)
are
\begin{align}
&\bigcup_{j=-\infty}^J \mathop{\bigcup_{(m,m')}}_{\neq (0,0)} \{M^{-j} \psi_m(\frac{x}{M^j}-k)
 \psi_{m'}(\frac{y}{M^j}-l), (k,l)\in \ZZ^2\} \nonumber \\
&\quad\bigcup\;\{M^{-J} \psi_0(\frac{x}{M^j}-k)
\psi_0(\frac{y}{M^j}-l), (k,l)\in \ZZ^2\}\\
&\bigcup_{j=-\infty}^J \mathop{\bigcup_{(m,m')}}_{\neq (0,0)} \{M^{-j} \psi_m^\HH(\frac{x}{M^j}-k)
 \psi_{m'}^\HH(\frac{y}{M^j}-l), (k,l)\in \ZZ^2\} \nonumber\\
&\quad \bigcup\;\{M^{-J} \psi_0^\HH(\frac{x}{M^j}-k)
 \psi_0^\HH(\frac{y}{M^j}-l), (k,l)\in \ZZ^2\} 
\end{align}
where $J\in\ZZ$ is the considered coarsest decomposition level.
A discrete implementation of these wavelet decompositions starts from 
level $j=1$ to $J\in \NN^*$.
As pointed out in the seminal works of Kingsbury and Selesnick,
it is however advantageous to add some pre- and post-processing
to this decomposition. The pre-processing aims at establishing the connection between the analog theoretical
framework and its discrete-time implementation whereas the post-processing is used to provide 
directional analysis features to the decomposition.
We will now revisit these problems in the context
of $M$-band decompositions.

The proposed $2D$ $M$-band dual-tree decomposition is illustrated in Fig. \ref{fig:Mbanddualtree}. For the sake of simplicity, 
only two levels of decomposition ($J=2$) are represented but this transform can
be implemented over further levels, the approximation coefficients being re-decomposed iteratively. 
For each of the two $M$-band decompositions,
we get $J\times M^{2}-J+1$ subbands. 
We observe that the $2D$ dual-tree decomposition can be divided into three steps which are detailed hereafter.

\subsubsection{Prefiltering} \label{se:prefilt}
The wavelet transform is a continuous-space formalism that we want to apply
to a ``discrete'' image.
We consider that the analog scene corresponds to the 2D field:
\begin{equation}
f(x,y)=\sum_{k,l}f[k,l]\;s(x-k,y-l)
\label{eq:cont_image}
\end{equation}
where $s$ is some interpolation function and $(f[k,l])_{(k,l)\in \ZZ^2}$
is the image sample sequence.
Let us project the image onto the approximation space
\begin{equation}
V_0 = \overline{\mathrm{Span}}\{\psi_0(x-k)\psi_0(y-l),(k,l)\in\ZZ^2\}.
\end{equation}
The projection of $f$ reads
\begin{equation}
P_{V_0}(f(x,y))=\sum_{k,l} c_{0,0,0}[k,l]\;\psi_0(x-k)\;\psi_0(y-l)
\end{equation}
where the approximation coefficients are
\begin{equation}
c_{0,0,0}[k,l]=\langle f(x,y),\psi_0(x-k)\;\psi_0(y-l) \rangle
\end{equation}
and $\langle \,,\,\rangle$ denotes the inner product of $\LL^2(\RR^2)$.
Using Eq.~\eqref{eq:cont_image} we obtain:
\begin{equation}
c_{0,0,0}[k,l] =\sum_{p,q} f[p,q]\;\gamma_{s,\Psi_{0,0}}(k-p,l-q)
\label{eq:prefilt}
\end{equation}
where $\Psi_{0,0}(x,y)=\psi_0(x)\psi_0(y)$ and $\gamma_{s,\Psi_{0,0}}$
is the cross-correlation function defined as
\begin{equation}
\gamma_{s,\Psi_{0,0}}(x,y) = \int_{-\infty}^\infty\!\int_{-\infty}^\infty s(u,v)\Psi_{0,0}(u-x,v-y)\;du\,dv.
\end{equation}
In the same way, we can project the analog image onto the
dual approximation space
\begin{equation}
V_0^\HH = \overline{\mathrm{Span}}\{\Psi_{0,0}^\HH(x-k,y-l),(k,l)\in\ZZ^2\}
\end{equation}
where $\Psi_{0,0}^\HH(x,y)=\psi_0^\HH(x)\psi_0^\HH(y)$.
We have then
\[
P_{V_0^\HH}(f(x,y))=\sum_{k,l}c_{0,0,0}^\HH[k,l]\;\Psi_{0,0}^\HH(x-k,y-l)
\]
where the dual  approximation coefficients are given by
\begin{equation}
c_{0,0,0}^\HH[k,l]=\sum_{p,q}f|p,q]\;\gamma_{s,\Psi_{0,0}^\HH}(k-p,l-q).
\label{eq:prefiltH}
\end{equation}
Obviously,
Eq.~\eqref{eq:prefilt} and \eqref{eq:prefiltH} can be interpreted 
as the use of two prefilters on the discrete image $(f[k,l])_{(k,l)\in\ZZ^2}$ before the dual-tree decomposition. The frequency response of these filters
are
\begin{align}
F_1(\omega_x,\omega_y) =&
 \sum_{p=-\infty}^\infty \sum_{q=-\infty}^\infty  
\widehat{s}(\omega_x+2p\pi,\omega_y+2q\pi) \nonumber \\
& \widehat{\psi}_0^*(\omega_x+2p\pi)\widehat{\psi}_0^*(\omega_y+2q\pi) \label{eq:expF1}\\
F_2(\omega_x,\omega_y) =&
 \sum_{p=-\infty}^\infty \sum_{q=-\infty}^\infty 
\widehat{s}(\omega_x+2p\pi,\omega_y+2q\pi) \nonumber \\
&  (\widehat{\psi}_0^\HH(\omega_x+2p\pi))^{*}(\widehat{\psi}_0^\HH(\omega_y+2q\pi))^*.
\label{eq:expF2}
\end{align}
By using Eq. \eqref{eq:linkpsi0Hpsi0}, it can be noticed that
\begin{equation}
F_2(\omega_x,\omega_y) = e^{\imath(d+1/2)(\omega_x+\omega_y)} F_1(\omega_x,\omega_y).
\label{eq:expF2F1}
\end{equation}
Different kinds of interpolation functions may be envisaged, in particular
separable functions of the form $s(x,y)=\chi(x)\chi(y)$. The two prefilters
are then separable with impulse responses  $(\gamma_{\chi,\psi_0}(p)\gamma_{\chi,\psi_0}(q))_{(p,q)\in \ZZ^2}$ and
$(\gamma_{\chi,\psi_0^\HH}(p)\gamma_{\chi,\psi_0^\HH}(q))_{(p,q)\in\ZZ^2}$,
respectively. A natural choice for $\chi$ is the Shannon-Nyquist interpolation function, $\chi(t)=\mathrm{sinc}(\pi t)$, which allows the ideal digital-to-analog conversion of a band-limited signal. We have then, for
$(\omega_x,\omega_y)\in [-\pi,\pi[^2$,
$F_1(\omega_x,\omega_y) = \widehat{\psi}_0^*(\omega_x)\widehat{\psi}_0^*(\omega_y)$.
Moreover, in the specific case when $\psi_0$ also corresponds to an ideal low-pass filter, that is $\psi_0(t) = \mathrm{sinc}(\pi t)$, the prefilter for the primal decomposition reduces to the identity ($F_1(\omega_x,\omega_y) = 1$)
whereas the prefilter for the dual decomposition is an half-integer shift with frequency response 
$F_2(\omega_x,\omega_y)=e^{\imath(d+1/2)(\omega_x+\omega_y)}$, for 
$(\omega_x,\omega_y)\in [-\pi,\pi[^2$.

\subsubsection{$M$-band wavelet decompositions}
The $M$-band multiresolution analysis of the first prefiltered image is
performed, resulting in coefficients 
\begin{equation}
c_{j,m,m'}[k,l]= \langle f(x,y),
\frac{1}{M^j}\psi_{m}(\frac{x}{M^j}-k) \psi_{m'}(\frac{y}{M^j}-l)\rangle
\end{equation}
where
($j\in \{1,\ldots,J\}$ and $(m,m')\neq(0,0)$) or ($j=J$ and $m=m'=0$). 
 In parallel, the dual decomposition of the second prefiltered image is computed, generating coefficients 
\begin{equation}
c_{j,m,m'}^\HH[k,l]= \langle f(x,y),
\frac{1}{M^j}\psi_{m}^\HH(\frac{x}{M^j}-k) \psi_{m'}^\HH(\frac{y}{M^j}-l)\rangle.
\end{equation}

\subsubsection{Direction extraction in the different subbands}
\label{sec:dirext}
\begin{figure} 
\centering
\input{figures/rot.pstex_t}
\caption{Direction selectivity in the 2D frequency plane
when $M=4$, $j=1$ and $(m,m')=(2,1)$. The two hachured areas, which are ``mixed'' in the original
$M$-band wavelet decomposition, can be separated by using tensor products of analytic/anti-analytic wavelets.}
\label{fig:rot2}
\end{figure}
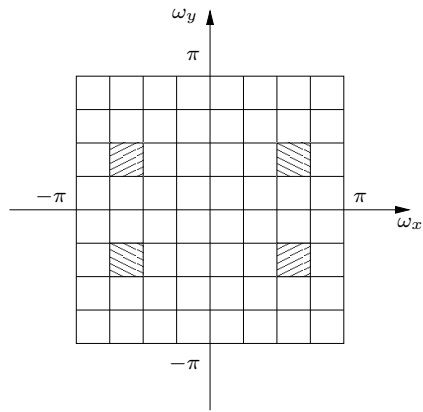

In order to better extract the local directions present in the image,
it is useful to introduce linear combinations of the primal
and dual subbands.
To do so, we define the analytic wavelets as
\begin{equation}
\psi_m^a(t) = \frac{1}{\sqrt{2}}(\psi_m(t) + \imath\; \psi_m^\HH(t)),
\qquad m \in \{0,\ldots,M-1\}
\label{eq:Analyticwav}
\end{equation}
and the anti-analytic wavelets as
\begin{equation}
\psi_m^{\bar{a}}(t) = \frac{1}{\sqrt{2}}(\psi_m(t) - \imath\; \psi_m^\HH(t)),
\qquad m \in \{0,\ldots,M-1\}.
\label{eq:Antianalyticwav}
\end{equation}
Let us now calculate the tensor product of two analytic wavelets
$\psi_m^a$ and $\psi_{m'}^a$. More precisely,
we are interested in the real part of this tensor product:
\begin{align}
\Psi_{m,m'}^a(x,y) &= \mathrm{Re}\{\psi_m^a(x)\psi_{m'}^a(y)\} \nonumber \\
&= \frac{1}{2}\big(\psi_m(x)\psi_{m'}(y)-\psi_m^\HH(x)\psi_{m'}^\HH(y)\big).
\label{eq:Analyticwav2}
\end{align}
For $(m,m')\in \{1,\ldots,M-1\}^2$, using Eq. (\ref{eq:Hilbertcond}), the Fourier transform of this
function is seen to be equal to
\begin{multline}
\widehat{\Psi}_{m,m'}^a(\omega_x,\omega_y)=\frac{1}{2}(1+\mathrm{sign}(\omega_x \, \omega_y))\widehat{\psi}_m(\omega_x)\widehat{\psi}_{m'}(\omega_y) \\
 = \left\{
\begin{array}{ll}
\widehat{\psi}_m(\omega_x)\widehat{\psi}_{m'}(\omega_y) & \mbox{if $\mathrm{sign}(\omega_x)=\mathrm{sign}(\omega_y)$},\\
0 & \mbox{if $\mathrm{sign}(\omega_x)\neq\mathrm{sign}(\omega_y)$}.
\end{array}
\right.
\label{eq:tensorielprod}
\end{multline}
As illustrated in Fig.~\ref{fig:rot2}, this function allows us to extract the
``directions'' falling in the first/third quarter of the frequency plane.

In the same way, the real part of the tensor product of an analytic wavelet and an anti-analytic one reads
\begin{equation}
\Psi_{m,m'}^{\bar{a}}(x,y) = \mathrm{Re}\{\psi_{m}^{a}(y)\psi_{m'}^{\bar{a}}(x)\} \label{eq:Antianalyticwav2}
\end{equation}
and, for $(m,m')\in \{1,\ldots,M-1\}^2$, its Fourier transform is
\begin{align}
\widehat{\Psi}_{m,m'}^{\bar{a}}(\omega_x,\omega_y) &= \\
& \left\{
\begin{array}{ll}
\widehat{\psi}_m(\omega_x)\widehat{\psi}_{m'}(\omega_y) & \mbox{if $\mathrm{sign}(\omega_x)\neq\mathrm{sign}(\omega_y)$},\\
0 & \mbox{if $\mathrm{sign}(\omega_x)=\mathrm{sign}(\omega_y)$}. \nonumber \\
\end{array}
\right.
\label{eq:tensorielprod2}
\end{align}
Fig. \ref{fig:rot2} shows that these functions allow us to select
frequency components which are localized in the second/fourth quarter
of the frequency plane. This yields ``opposite'' directions
to those obtained with $\Psi_{m,m'}^a$.

At a given resolution level $j$, for each subband $(m,m')$ with $m\neq 0$
and $m'\neq 0$, the directional analysis is achieved by computing
the coefficients
\begin{align}
d_{j,m,m'}[k,l] = &\langle f(x,y),
\frac{1}{M^j}\Psi_{m,m'}^{\bar{a}}(\frac{x}{M^j}-k,\frac{y}{M^j}-l)\rangle\\
d_{j,m,m'}^\HH[k,l] = &\langle f(x,y),
\frac{1}{M^j}\Psi_{m,m'}^a(\frac{x}{M^j}-k,\frac{y}{M^j}-l)\rangle.
\end{align}
According to Eqs.~\eqref{eq:Analyticwav}, \eqref{eq:Analyticwav2},
\eqref{eq:Antianalyticwav} and \eqref{eq:Antianalyticwav2}, we have for all $(m,m')\in \{1,\ldots,M-1\}^2$,
\begin{align}
\label{eq:rotation}
d_{j,m,m'}[k,l]&=\frac{1}{\sqrt{2}}(c_{j,m,m'}[k,l]+c^\HH_{j,m,m'}[k,l])\\
d_{j,m,m'}^\HH[k,l]&=\frac{1}{\sqrt{2}}(c_{j,m,m'}[k,l]-c^\HH_{j,m,m'}[k,l])
\end{align}
which amounts to applying a simple $2\times 2$ isometry to the $M$-band
wavelet coefficients.
Note that Relations~\eqref{eq:tensorielprod} and \eqref{eq:tensorielprod2} are not valid 
for horizontal or vertical low-pass subbands such that $m=0$ or $m'=0$.
The corresponding coefficients are left unchanged by setting
$d_{j,m,m'}[k,l] = c_{j,m,m'}[k,l]$ and $d_{j,m,m'}^\HH[k,l] = c_{j,m,m'}^\HH[k,l]$.

\begin{figure} 
\begin{center}
\begin{tabular}{c c}
\includegraphics[width=8.5cm]{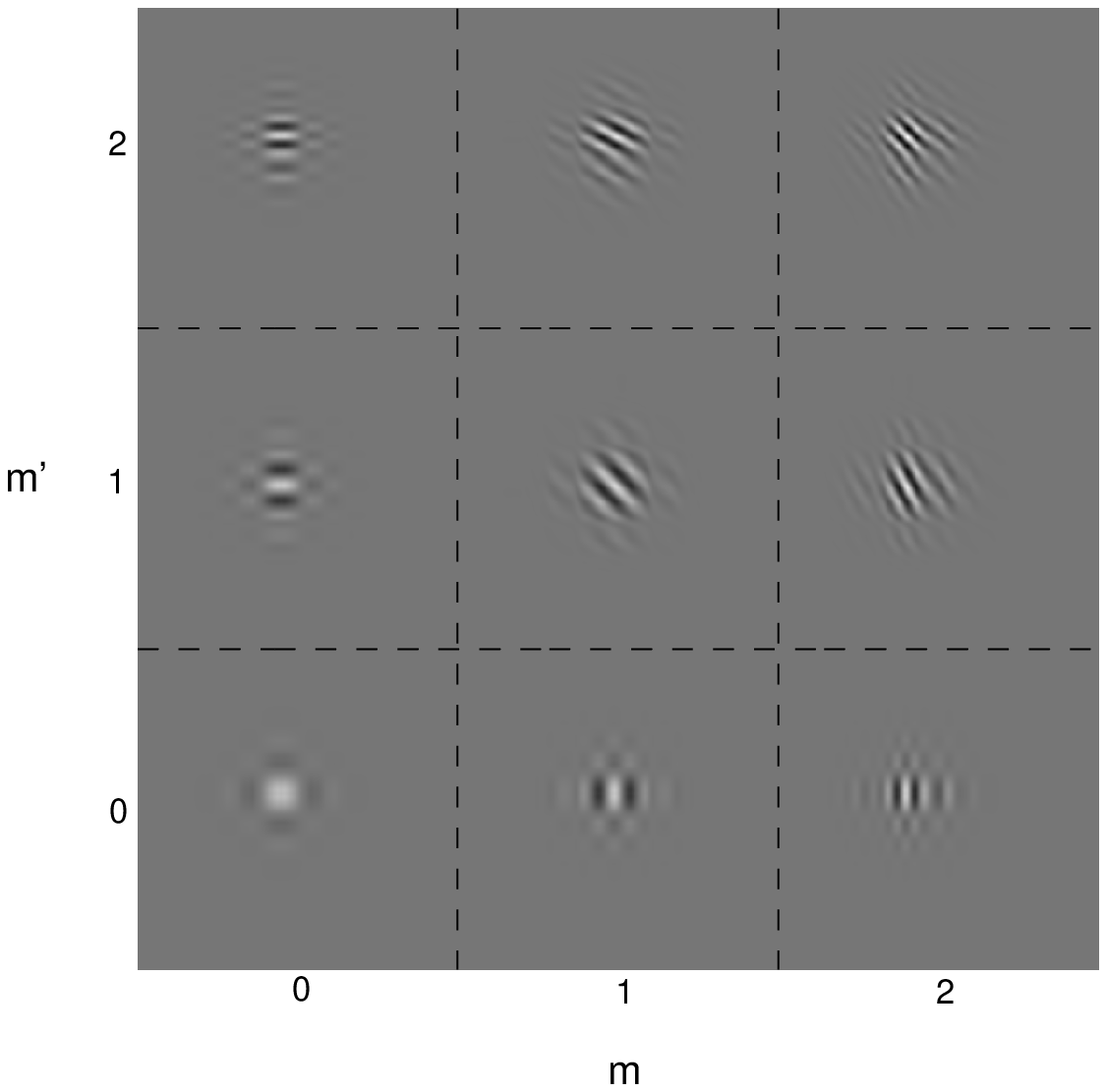} & \includegraphics[width=8.5cm]{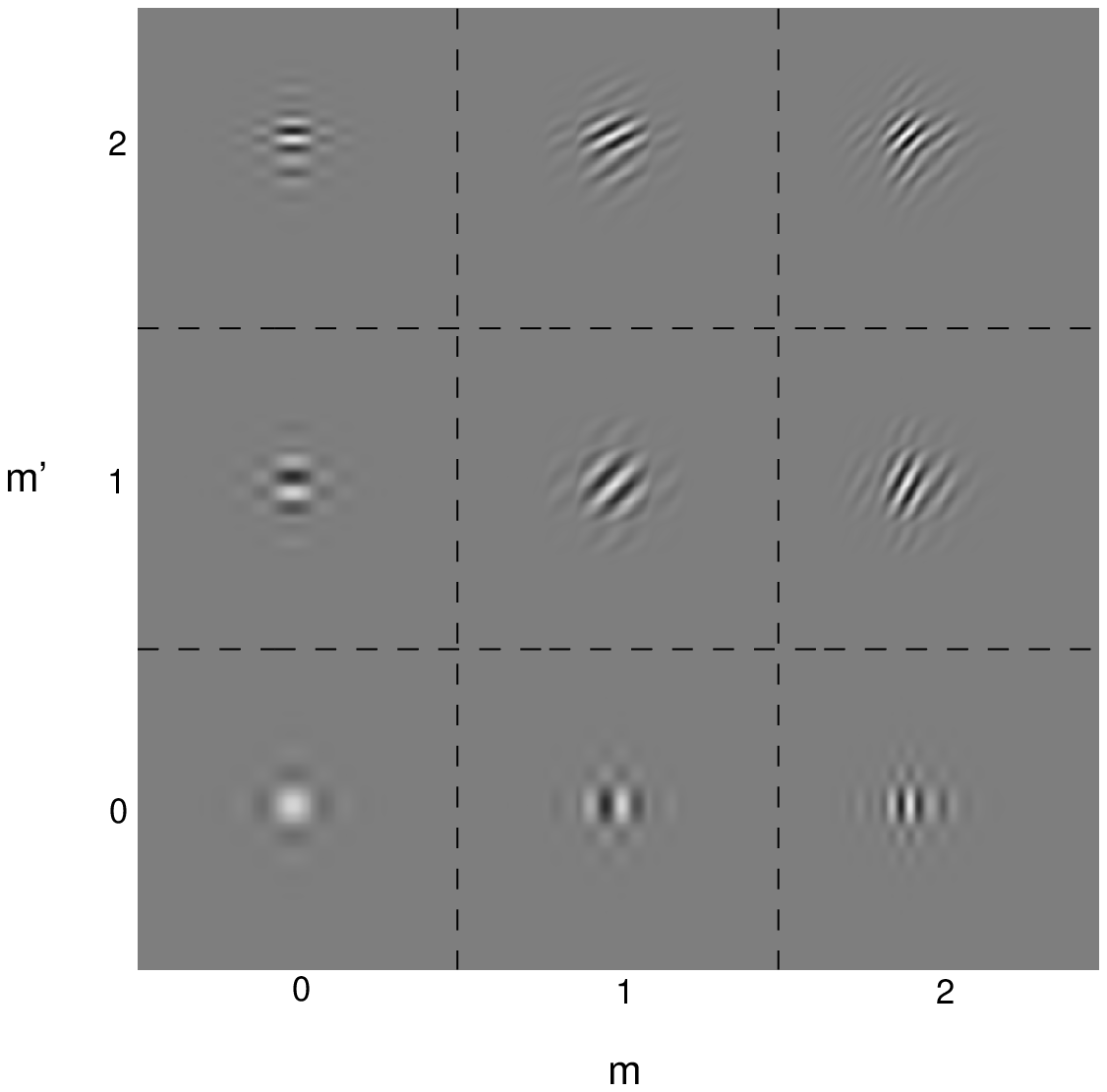} \\
\end{tabular}
\end{center}
\caption{Functions $(\Psi_{m,m'}^{a}(x,y))_{0\le m,m' \le 2}$
(on the left)
and $(\Psi_{m,m'}^{\bar{a}}(x,y))_{0\le m,m' \le 2}$ (on the right).
These functions are derived from 3-band Meyer's wavelets and their associated dual functions.}
\label{fig:wav}
\end{figure}

To illustrate the  improved directional analysis provided by
the proposed decompositions, the basis functions used in a
3-band dual-tree structure are shown in
Fig. \ref{fig:wav}.

\subsection{Reconstruction}
Let us denote by $\mathbf{f}\in \ell^2(\ZZ^2)$ the vector of image samples 
where $\ell^2(\ZZ^2)$ is the space of finite-energy 2D discrete fields. Besides, we denote by $\mathbf{c}$ the vector of coefficients generated by the primal  $M$-band decomposition and by $\mathbf{c^\HH}$ the vector of coefficients produced by the dual one. 
These vectors consist of $MJ-J+1$ sequences each belonging to $\ell^2(\ZZ^2)$.
The linear combination of the subbands described in
Section~\ref{sec:dirext}  can be omitted in the subsequent analysis since we have seen
that this post-processing reduces to a trivial $2\times 2$ orthogonal 
transform. The global decomposition operator (including decomposition steps 1 and 2) is
\begin{align}
\mathbf{D}\ :
\mathbf{f} &\mapsto \begin{pmatrix}
\mathbf{c}\\
\mathbf{c^\HH}
\end{pmatrix} =  \begin{pmatrix}
\mathbf{D_1}  \mathbf{f}\\
\mathbf{D_2} \mathbf{f}
\end{pmatrix}
\end{align}
where $\mathbf{D_1}=\mathbf{U_1F_1}$ and $\mathbf{D_2}=\mathbf{U_2F_2}$,
$\mathbf{F_1}$ and $\mathbf{F_2}$ being the prefiltering operations
described in Section~\ref{se:prefilt} and $\mathbf{U_1}$ and $\mathbf{U_2}$ being the two considered orthogonal $M$-band wavelet decompositions.
We have then the following result whose proof is provided in Appendix \ref{ap:fram}:
\begin{prop} \label{prop:fram}
Provided that there exist positive constants 
$A_s$, $B_s$, $C_s$ and $A_{\psi_0}$
such that, for (almost) all $(\omega_x,\omega_y)\in [-\pi,\pi[^2$,
\begin{align}
&\qquad \qquad  A_s \leq |\widehat{s}(\omega_x,\omega_y)| \leq B_s ,\quad |\widehat{\psi}_0(\omega_x)| \geq A_{\psi_0}
\label{eq:bounds}\\
&\sum_{(p,q)\neq(0,0)}|\widehat{s}(\omega_x+2p\pi,\omega_y+2q\pi)|^2\leq C_s < A_s^2A_{\psi_0}^4
\label{eq:boundpsi0}
\end{align}
$\mathbf{D}$ is a frame operator.
The ``dual'' frame reconstruction operator
is given by\footnote{Here ``dual'' is meant in the sense
of the frame theory \cite{Daubechies_I_1992_book_ten_lw} which is different from the sense given
in the rest of the paper.}
\begin{equation}
\mathbf{f} = (\mathbf{F_1}^{\dagger}\mathbf{F_1}+\mathbf{F_2}^{\dagger}\mathbf{F_2})^{-1} \; (\mathbf{F_1}^{\dagger}\mathbf{U_1}^{-1}\mathbf{c} + \mathbf{F_2}^{\dagger}\mathbf{U_2}^{-1}\mathbf{c^H})
\label{eq:pseudoinverse}
\end{equation}
where $\mathbf{T}^{\dagger}$ designates the adjoint of an operator $\mathbf{T}$.
\end{prop}

A particular case of interest is when $\{s(x-k,y-l), (k,l)\in \ZZ^2\}$
is an orthonormal family of $\LL^2(\RR^2)$. We then have 
$\sum_{p,q} |\widehat{s}(\omega_x+2p\pi,\omega_y+2q\pi)|^2 = 1$ and consequently
we can choose $B_s = 1$. The lower bounds $A_s$ and $A_{\psi_0}$ prevent $\widehat{s}$ and $\widehat{\psi}_0$ from vanishing for low frequencies whereas Eq. \eqref{eq:bounds}
controls the amount of energy of $\widehat{s}$ out of the frequency band
$[-1/2,1/2[^2$. Note that the assumptions on $s$ are obviously satisfied by the Shannon-Nyquist interpolation function.

Although other reconstructions of $\mathbf{f}$ from $(\mathbf{c},\mathbf{c^\HH})$ could be envisaged, Formula~\eqref{eq:pseudoinverse} minimizes the impact of 
possible errors in the computation of the wavelet coefficients. 
For example, these errors may arise in the estimation procedures 
when a denoising application is considered.
Finally, it is worth pointing out that Eq.~\eqref{eq:pseudoinverse} is not difficult to implement since
$\mathbf{U_1}^{-1}$ and $\mathbf{U_2}^{-1}$ are the inverse $M$-band wavelet
transforms and $\mathbf{F_1}^{\dagger}$, $\mathbf{F_2}^{\dagger}$
and $(\mathbf{F_1}^{\dagger}\mathbf{F_1}+\mathbf{F_2}^{\dagger}\mathbf{F_2})^{-1}$ correspond to filtering with frequency responses 
$F_1^*(\omega_x,\omega_y)$, $F_2^*(\omega_x,\omega_y)$
and $(|F_1(\omega_x,\omega_y)|^2+|F_2(\omega_x,\omega_y)|^2)^{-1}$, respectively.

\section{Implementation and design issues} \label{se:implement}
\subsection{$M$-band wavelet and filter bank families}
\begin{figure}
\begin{center}
\begin{tabular}{c c}
\includegraphics[width=8cm]{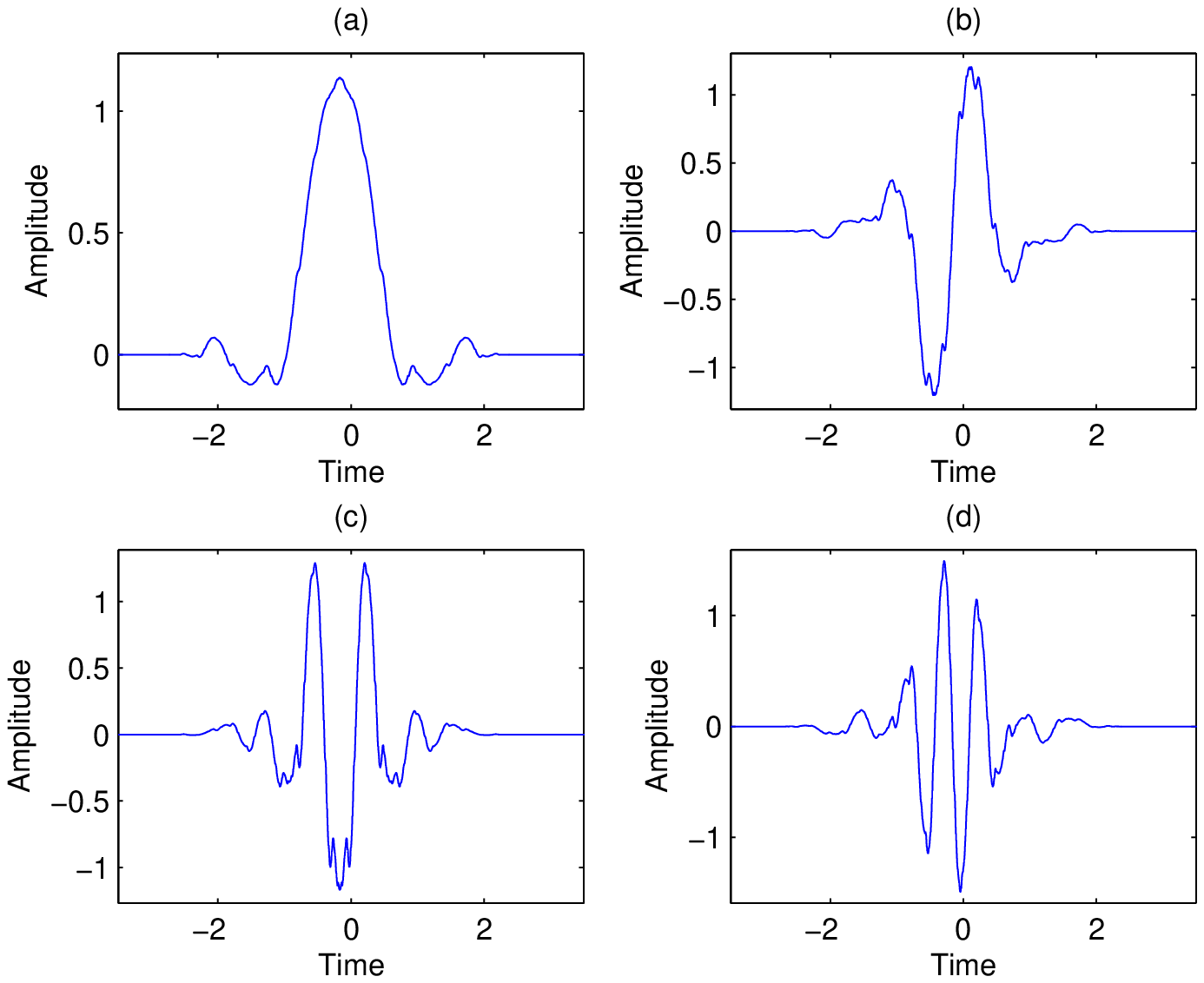}&
\includegraphics[width=8cm]{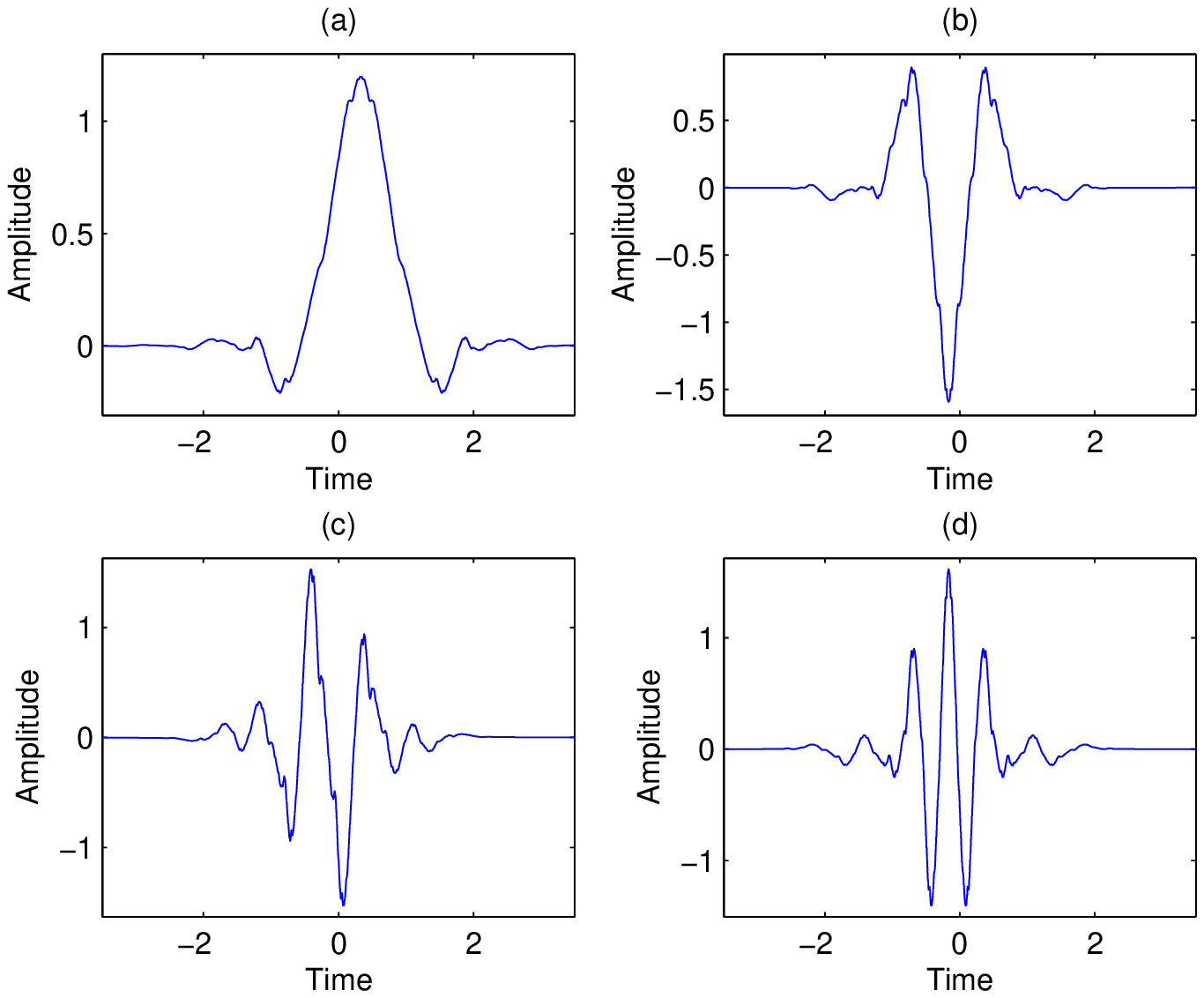}\\
(I) & (II)\\
\end{tabular}
\end{center}
\caption{(I): (a) Scaling function $\psi_0$ and (b) wavelet $\psi_1$, (c)~wavelet $\psi_2$, (d) wavelet~$\psi_3$ and (II): (a) Scaling function $\psi_0^{\mathrm{H}}$ and (b) wavelet $\psi_1^{\mathrm{H}}$, (c)~wavelet $\psi_2^{\mathrm{H}}$, (d) wavelet $\psi_3^{\mathrm{H}}$\label{fig:wavelets1D} with filters derived from  \cite{Alkin_O_1995_j-ieee-tsp_design_embclpprp}.
These functions have been generated using the scaling equations
\eqref{eq:twoscale} and \eqref{eq:twoscaled} in the frequency
domain.
}
\end{figure}

In our experiments, the advantage of the dual-tree decomposition has been tested over several classical dyadic orthonormal wavelet bases. Since we are interested in its $M$-band generalization,  
several other $M$-band filter banks decompositions have been considered, including both $M$-band wavelets and lapped transforms 
(we refer to  \cite{Oraintara_S_2001_tsp_lat_srplpfmbosw,Chen_Y_2003_j-ieee-tcas2_mchannel_lfprfbrmwt} for more details on filter banks regularity):
\begin{itemize}
\item Primal wavelets with compact support: the first  example consists in four finite impulse response (FIR) 21-tap filters  (denoted as AC in \cite{Alkin_O_1995_j-ieee-tsp_design_embclpprp}), generating regular, orthonormal and symmetric basis functions.
The scaling function and the wavelets associated to the dual $4$-band filter bank are represented in Fig.~\ref{fig:wavelets1D}.
 We observe that the constructed dual wavelets possess regularity and
satisfy the  symmetry properties stated in Proposition \ref{prop:sym}. We also have constructed and tested dual wavelets from  standard symmlets as well as a $4$-channel modulated lapped transform \cite{Malvar_H_1992_book_signal_plt}.

\item Primal wavelets without compact support: we have constructed $M$-band generalization of Meyer's wavelets. The corresponding filters possess a good
frequency selectivity.
To implement these filters, we have used a  method similar to that developped in \cite{Tennant_B_2003_solution_omcbwcp}. 
Taking the same wavelet family with a different number of bands helps
in providing fair assessment on the benefits of using more channels.
\end{itemize}

\subsection{Frequency-domain implementation}
Two solutions are possible to implement a wavelet decomposition:
a time-domain  or a frequency-domain approach.
The first one is probably the most popular for classical wavelet
decompositions when wavelets with compact support are used.
Sometimes however, especially for wavelets having an infinite support (for instance
orthonormal spline wavelets), a frequency-domain implementation is often preferable,
 taking advantage of FFT algorithms \cite{Nicolier_F_2002_j-elec-im_dis_wtifdms} (see also
\cite{Rioul_O_1992_tit_fas_adcwt} for a thorough discussion of these problems). In particular, FFTs are used to compute Fractional Spline Wavelet Transform \cite{Blu_T_2000_icassp_fra_swtdi} and also to implement steerable pyramids \cite{Simoncelli_E_1995_icip_ste_pfamdc}.
In the case of dual-tree decompositions, we have noticed in Section \ref{se:compactsup}
that, when the primal wavelets are compactly supported, the dual ones are not. If a time-domain implementation is chosen, it then becomes  necessary
to approximate the infinite impulse responses of the dual filter bank by
finite sequences satisfying  constraints related to the para-unitarity conditions, symmetry, number of vanishing moments, etc. The resulting optimal
design problem may become involved and, for a good approximation
of the ideal dual responses, it may happen that
the obtained solutions only approximately
satisfy  the para-unitarity conditions which correspond to non-convex constraints. In spite of these difficulties, such an approach was followed in \cite{Selesnick_I_2002_j-ieee-tsp_design_ahtpwb} 
which is approximate in the sense of the Hilbert transform and symmetry
and in one of our previous work \cite{Chaux_C_2004_p-eusipco_hilbert_pmbowb}.
For the simulations in this paper, frequency-domain implementations have been adopted. They may
provide better numerical solutions in the context of dual-tree decompositions.
In this case, both convolutions and decimations are performed in the frequency domain.

\section{Application to denoising}
\label{sec:denoise}
The $2$-band multidimensionnal dual-tree complex wavelet transform  has already been proved to be useful in denoising problems, in particular for video processing\cite{Fernandes_F_2003_j-sp_com_wtaf} or satellite imaging \cite{Jalobeanu_A_2003_ijcv_sat_idcwp}.
In this part, we show that $M$-band dual-tree wavelet transforms also demonstrate good performances in image denoising and outperform existing 
methods such as those relying on classical $M$-band wavelet transforms ($M\geq2$) or even $2$-band dual-tree wavelet transforms. We will be mainly interested
in applications involving images containing directional information and texture-like behavior such as seismic images.

\subsection{Denoising problem}
In this part, we will consider the estimation of an image $s$, corrupted by an additive zero-mean white Gaussian noise $b$ with power spectrum density $\sigma^2$. The observed image $f(x,y)$ is therefore given by: $f(x,y)=s(x,y)+b(x,y)$.
We will denote by $(b_{j,m,m'}[k,l])_{k,l}$ the coefficients resulting from a 2D $M$-band wavelet decomposition of the noise
in a given subband $(j,m,m')$. The associated wavelet coefficients of the dual decompositions are denoted by
$(b^\HH_{j,m,m'}[k,l])_{k,l}$. These sequences are white zero-mean Gaussian with variance $\sigma^2$.
Besides, we have for all $(k,l)\in\ZZ^2$,

\begin{multline}
\E\{b_{j,m,m'}[k,l]b^{\HH}_{j,m,m'}[k,l]\}=\\
\int_{\RR^4} \E\{b(x,y)b(x',y')\}\frac{1}{M^{j}}\psi_m(\frac{x}{M^j}-k)\psi_{m'}(\frac{y}{M^j}-l)\\\frac{1}{M^{j}}\psi_m^\HH(\frac{x'}{M^j}-k)\psi^\HH_{m'}(\frac{y'}{M^j}-l)\;dxdydx'dy'
\end{multline}
where $\E\{b(x,y)b(x',y')\}=\sigma^2\;\delta(x-x')\delta(y-y')$ ($\delta$ is the Dirac distribution). After some straightforward calculations
when $m\neq 0$ or $m'\neq 0$, this yields
\begin{equation}
\E\{b_{j,m,m'}[k,l]b_{j,m,m'}^\HH[k,l]\} = 0.
\label{eq:internoise}
\end{equation}
It is deduced that, when $m\neq0$ or $m'\neq0$, 
the Gaussian vector $(b_{j,m,m'}[k,l]\quad b_{j,m,m'}^\HH[k,l])^T$ has independent components.

The variance of the noise may be unknown. In such a case, we use a robust estimator $\hat{\sigma}$ of $\sigma$ which is computed from the wavelets coefficients at scale $j=1$ in a high-pass subband (see \cite[p. 447]{Mallat_S_1998_book_wav_tsp}):
\begin{equation}
 \hat{\sigma}=\frac{1}{0.6745}\;\text{median}[(|c_{1,M-1,M-1}[k,l]|)_{(k,l)}].
\label{eq:med_est}
\end{equation}

\subsection{Thresholding}
Various thresholding techniques have been applied on the wavelet
coefficients of the observed image $f$.
Although many choice of estimators can be envisaged, we have
studied the following ones:
\begin{itemize}
\item Visushrink (see \cite{Donoho_D_1994_j-biometrika_ideal_saws}) 
defined by the ``universal'' hard threshold $T=\sigma \; \sqrt{2\;\ln(N)}$, $N$ being the number of pixels of the original image.
\item Hybrid SUREshrink  \cite{Donoho_D_1995_jasa_ada_usws,Krim_H_1999_tit_den_bsr}.
This subband-adaptive threshold technique relies on Stein's Unbiased Risk Estimate and uses a soft thresholding. As a result, if the signal to noise ratio is very small, the SURE estimate may become unreliable. If such a situation is detected, a universal threshold is used.
\item Cai and Silverman estimator \cite{Cai_T_2001_sankhya_inc_incwe}. 
This block thresholding approach exploits correlations between neighboring coefficients. 
In our work, we use a variant of the NeighBlock method.
\item Bivariate Shrinkage \cite{Sendur_L_2002_j-ieee-spl_bivariate_slve}.
This method  exploits the interscale dependencies i.e. relations between the coefficients and their parents.

\end{itemize}

\subsection{Mesures of performance}
Let $N$ be the number of points in the observed image $f$,
$\sigma_s$ the standard deviation of $s$. We define two signal-to-noise ratios, denoted by $\textrm{SNR}$, as:
\begin{align}
&\textrm{SNR}_\textrm{initial}=10\;\log_{10}\left(\frac{\sigma_s^2\;N}{\|s-f\|^2}\right) \nonumber \\
&\textrm{SNR}_\textrm{final}=10\;\log_{10}\left(\frac{\sigma_s^2\;N}{\|s-\hat{s}\|^2}\right)
\end{align}
where $\hat{s}$ is the estimated image.

Visual comparisons are provided as well, since SNR does not always faithfully accounts for image quality, especially in highly structured areas (textures, edges,...)

\subsection{Experimental results}
Tests have been carried out on a variety of images corrupted by an additive zero-mean white Gaussian noise. We have considered two possible situations : first, when the noise variance is  known and second, when it  is not. In the latter case, the noise variance is estimated with the robust median estimator as defined in Eq. (\ref{eq:med_est}).
The noisy image is decomposed via an $M$-band
DWT or an $M$-band Dual-Tree Transform (DTT) in the $2$, $3$ and $4$-band cases. For each decomposition, the number of decomposition levels is fixed so as to get approximation images having roughly
the same size at the coarsest resolution. This means that
2-band decompositions are carried out over 4 resolution,
whereas 3 or 4-band decompositions are performed over 2 resolution levels.
Under these conditions, the computational costs of the different
$M$-band decompositions are comparable.
Different wavelet families have been tested, the provided results corresponding
to the use of Meyer's wavelets \cite{Tennant_B_2003_solution_omcbwcp}.
For various noise levels, the values of the SNR's are obtained from 
a Monte Carlo study over ten noise realizations.

Since we address more specifically the ability of the $M$-band DTT to preserve features in specific directions, comparisons are made on the following 
three images containing rich directional contents: a high frequency textured image, the standard Barbara image and a set of 2D seismic data with oriented patterns. 

\begin{figure} 
\begin{center}
\begin{tabular}{c c}
\includegraphics[width=4cm]{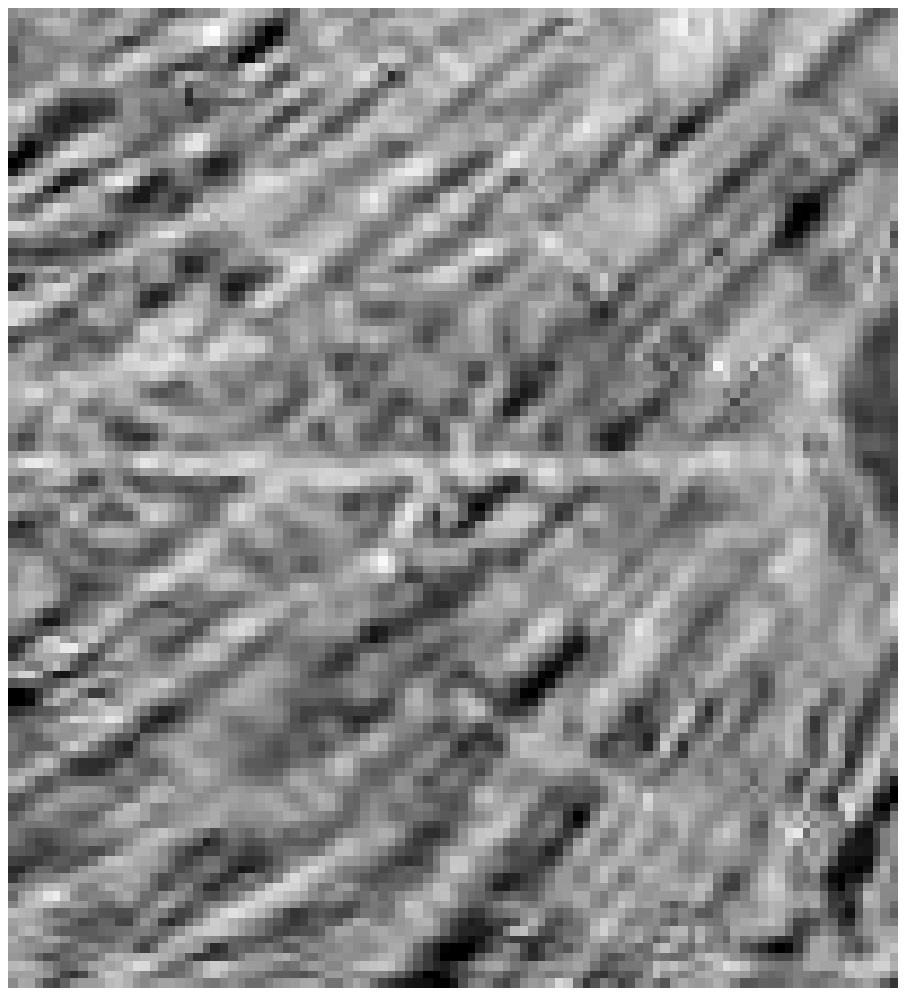}&
\includegraphics[width=4cm]{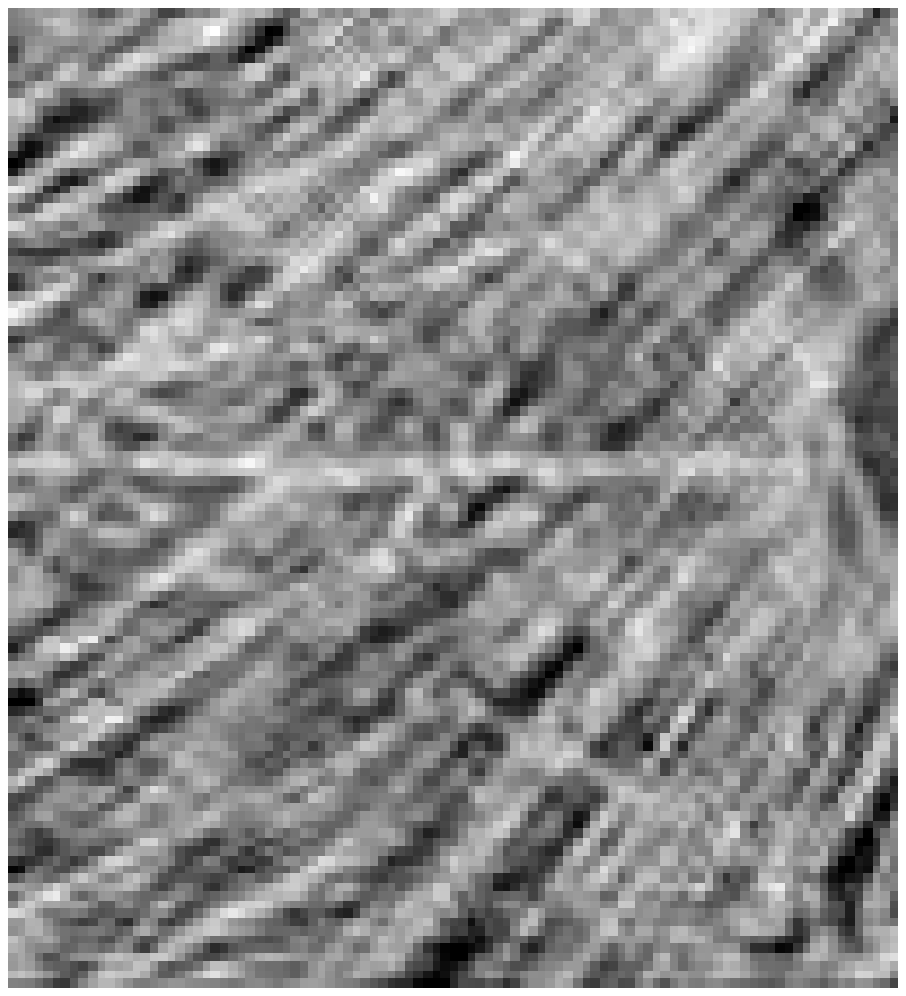}\\
(a) & (b) \\

\includegraphics[width=4cm]{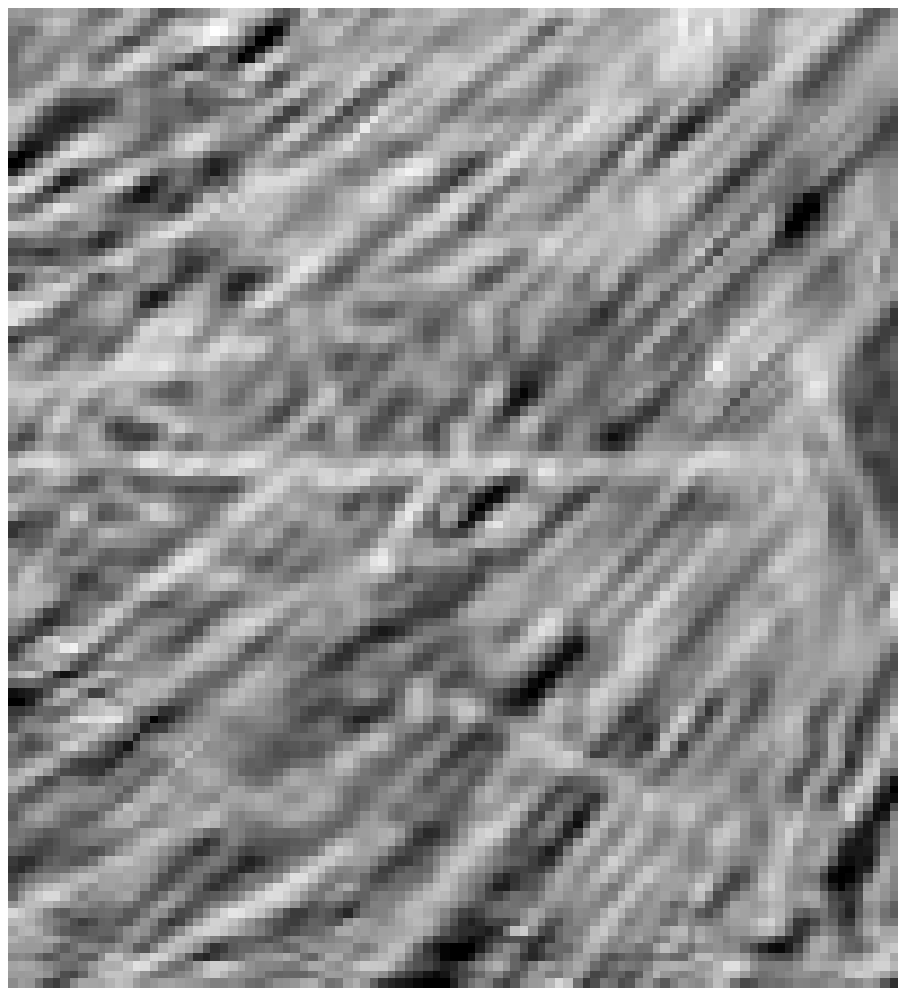}&
\includegraphics[width=4cm]{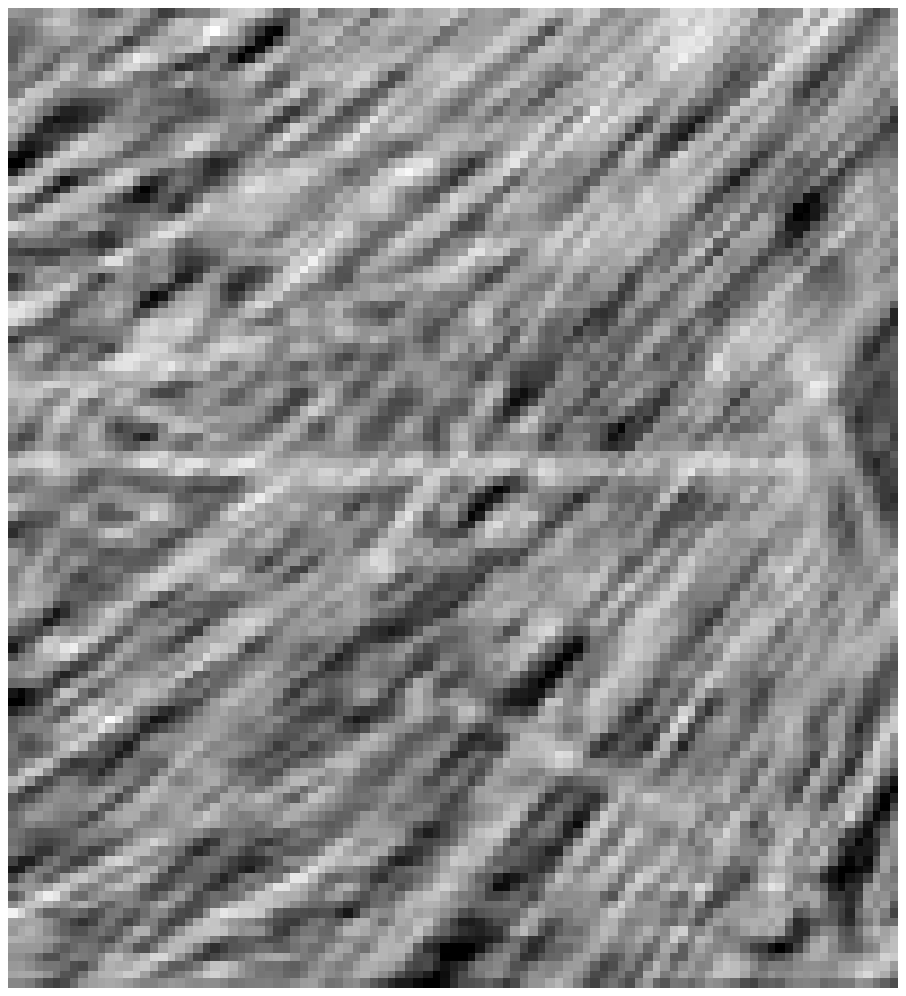}\\
(c) & (d) \\
\end{tabular}
\end{center}
\caption{Denoising results for a cropped version of the texture using Bivariate Shrinkage and: (a) DWT $M=2$; (b) DWT $M=4$; (c) DTT $M=2$; (d) DTT $M=4$. {\label{fig:zoomtextu}}}
\end{figure}

\begin{figure} 
\begin{center}
\begin{tabular}{c c}
\includegraphics[width=4cm]{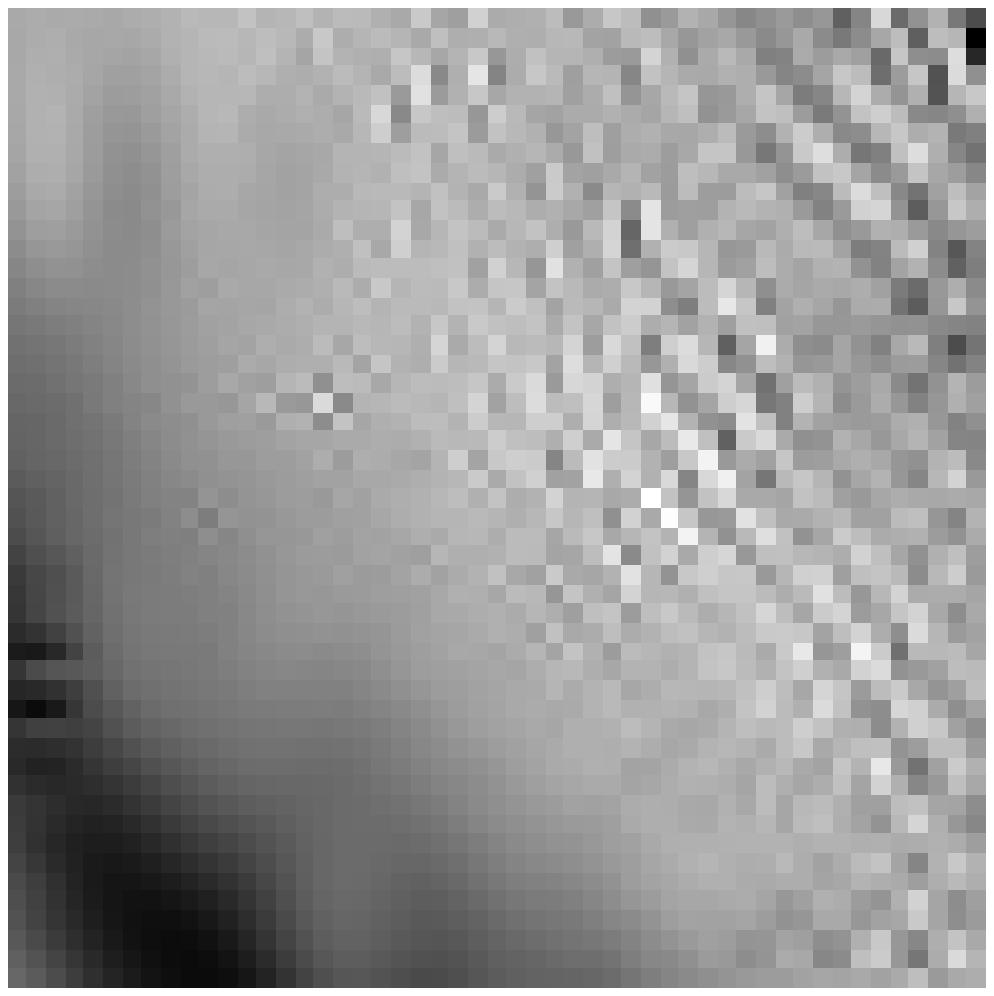}&
\includegraphics[width=4cm]{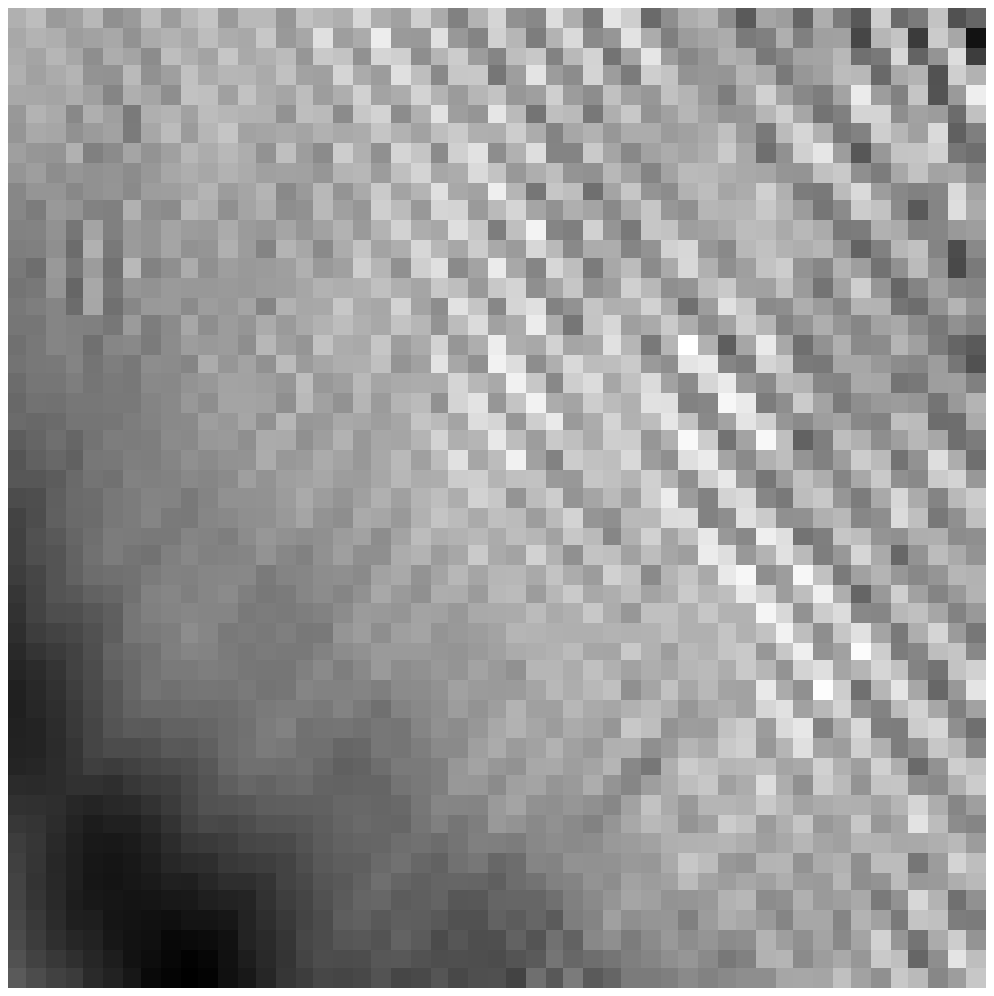}\\
(a) & (b) \\

\includegraphics[width=4cm]{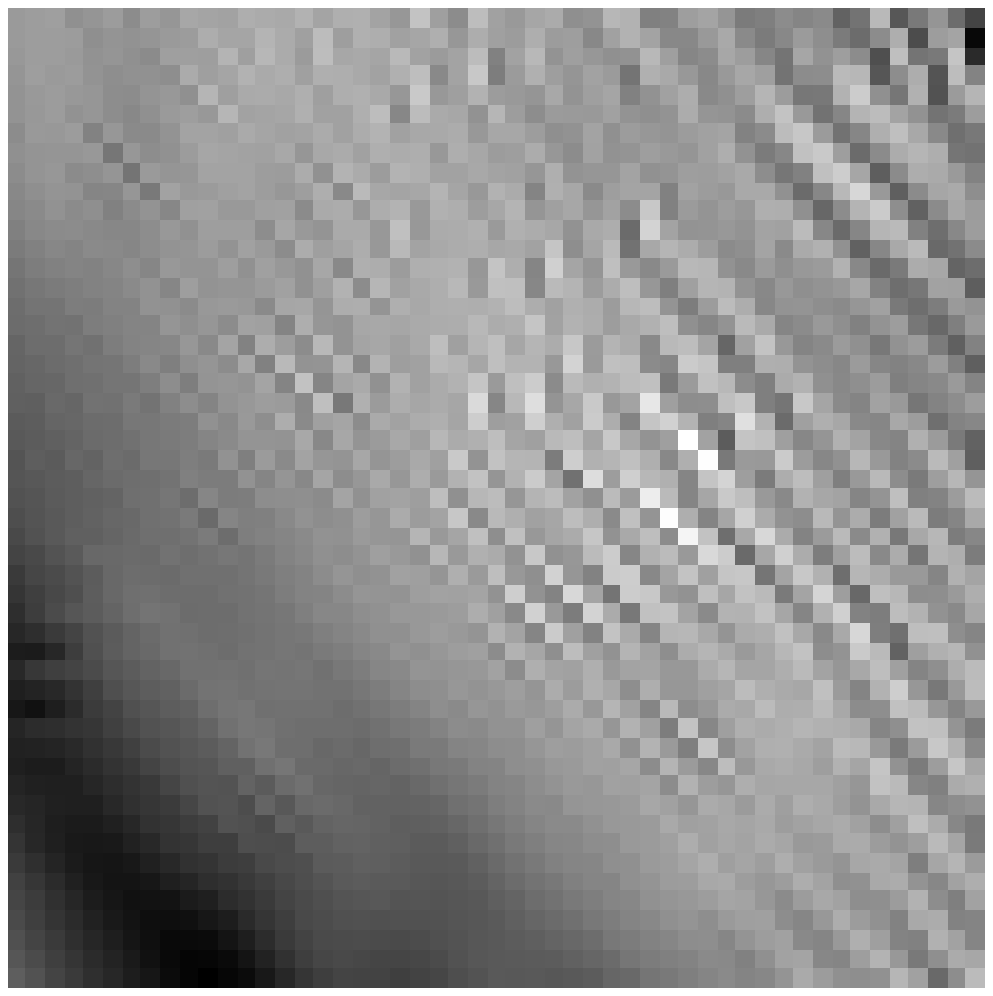}&
\includegraphics[width=4cm]{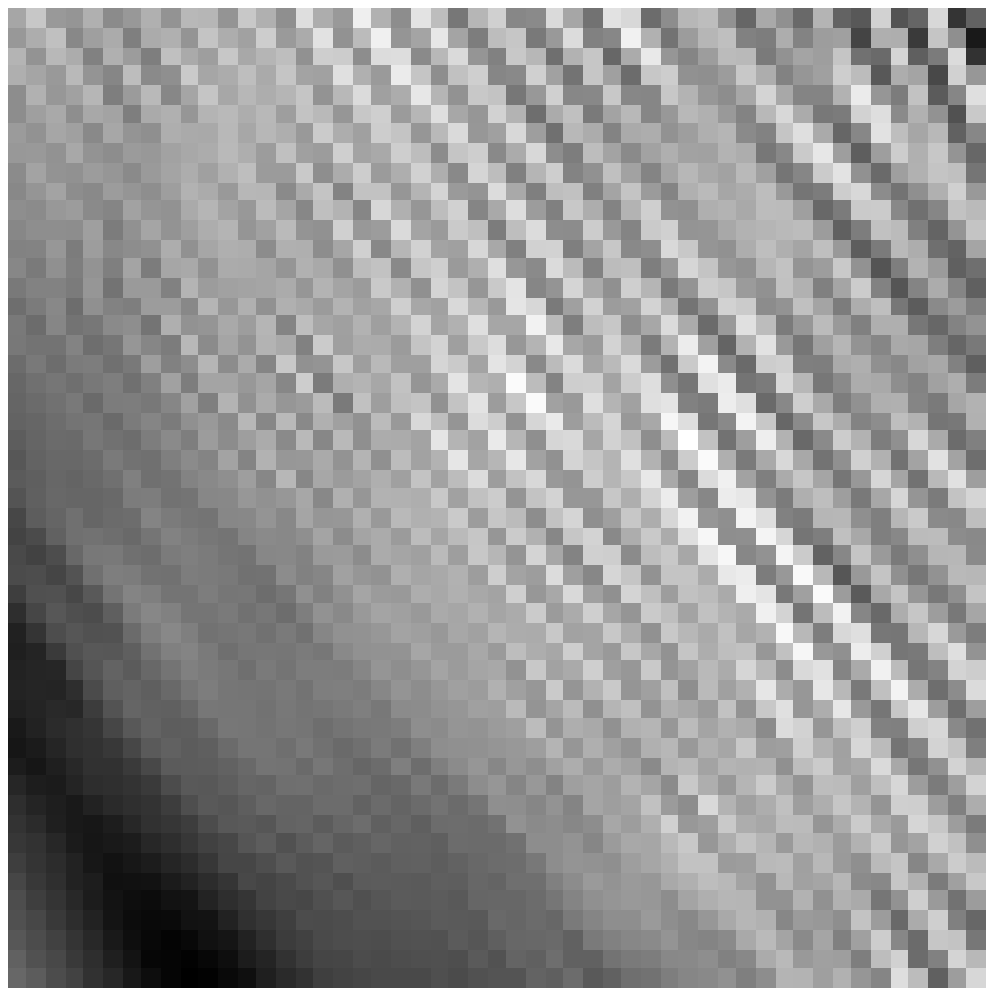}\\
(c) & (d) \\
\end{tabular}
\end{center}
\caption{Denoising results for a cropped version of ``Barbara'' using Bivariate Shrinkage and: (a) DWT $M=2$; (b) DWT $M=4$; (c) DTT $M=2$; (d) DTT $M=4$. {\label{fig:zoombarb1}}}
\end{figure}

\begin{figure} 
\begin{center}
\begin{tabular}{c c}
\includegraphics[width=8cm,height=4cm]{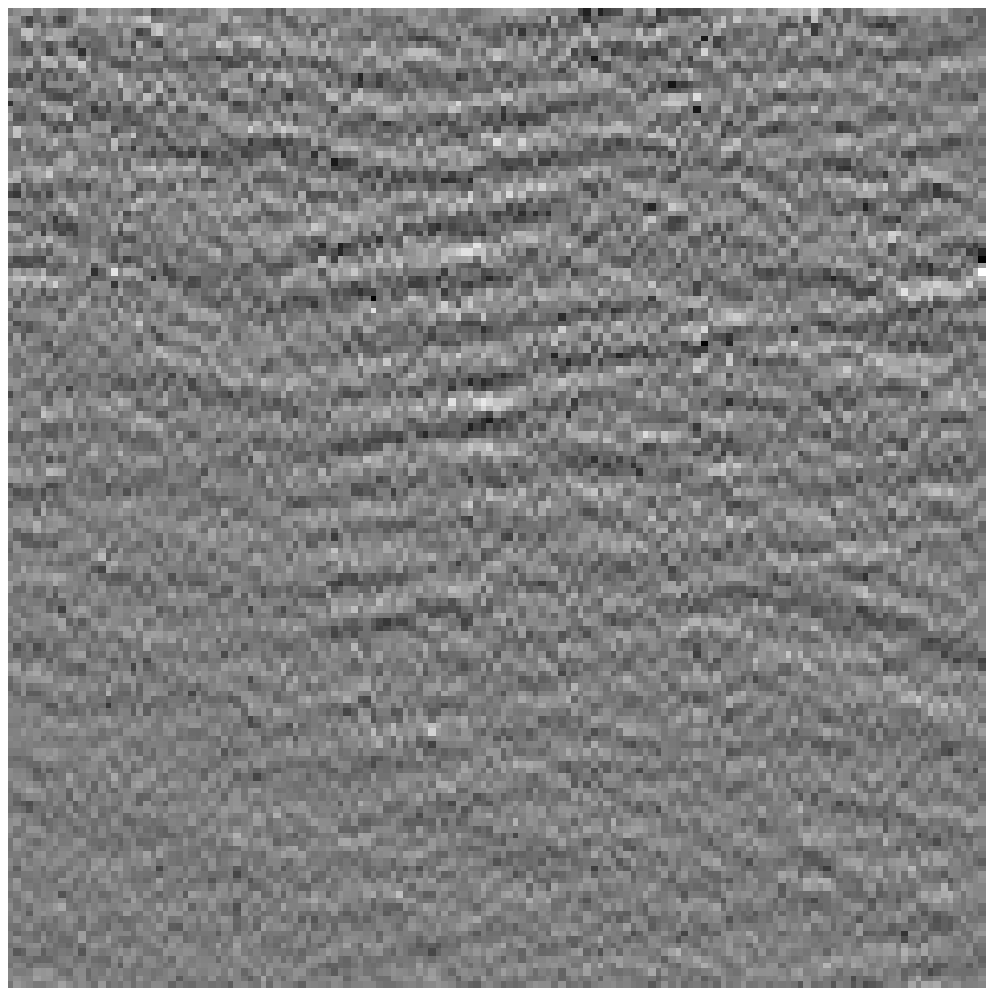}&
\includegraphics[width=8cm,height=4cm]{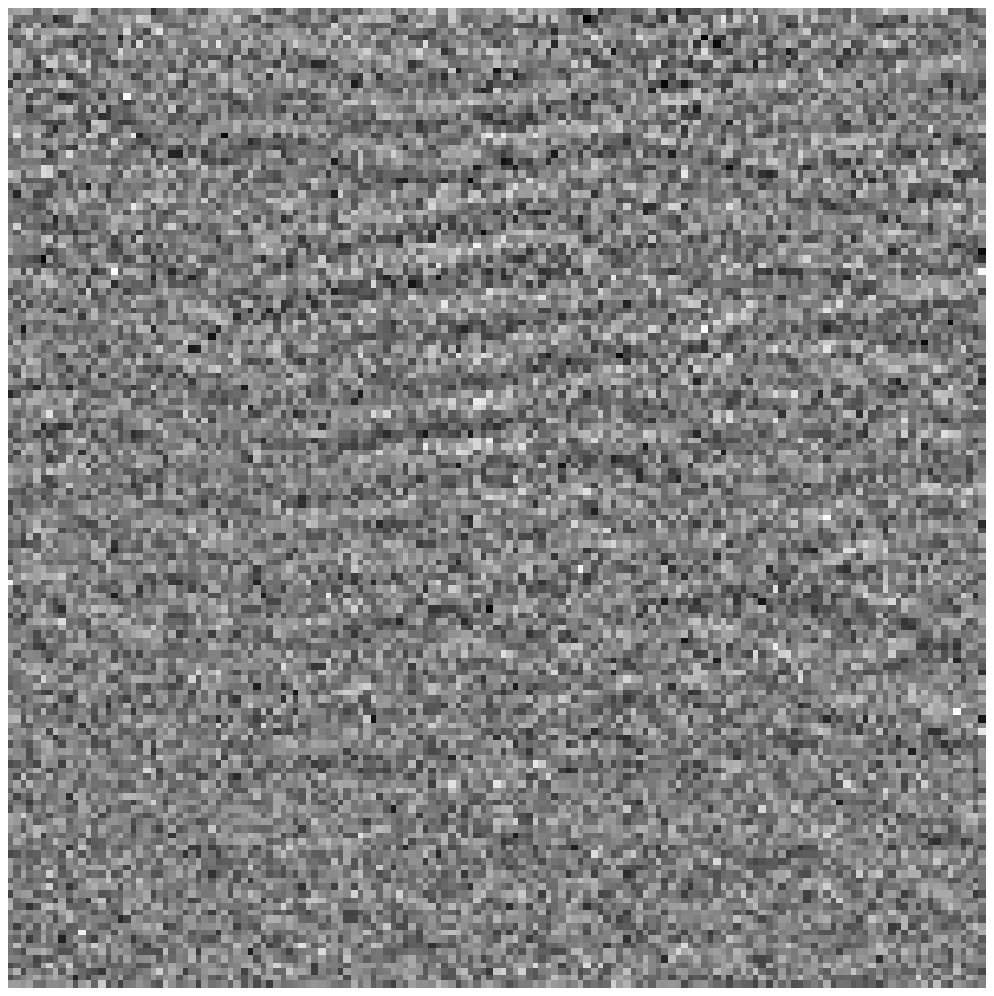} \\
(a) & (b)  \\

\includegraphics[width=8cm,height=4cm]{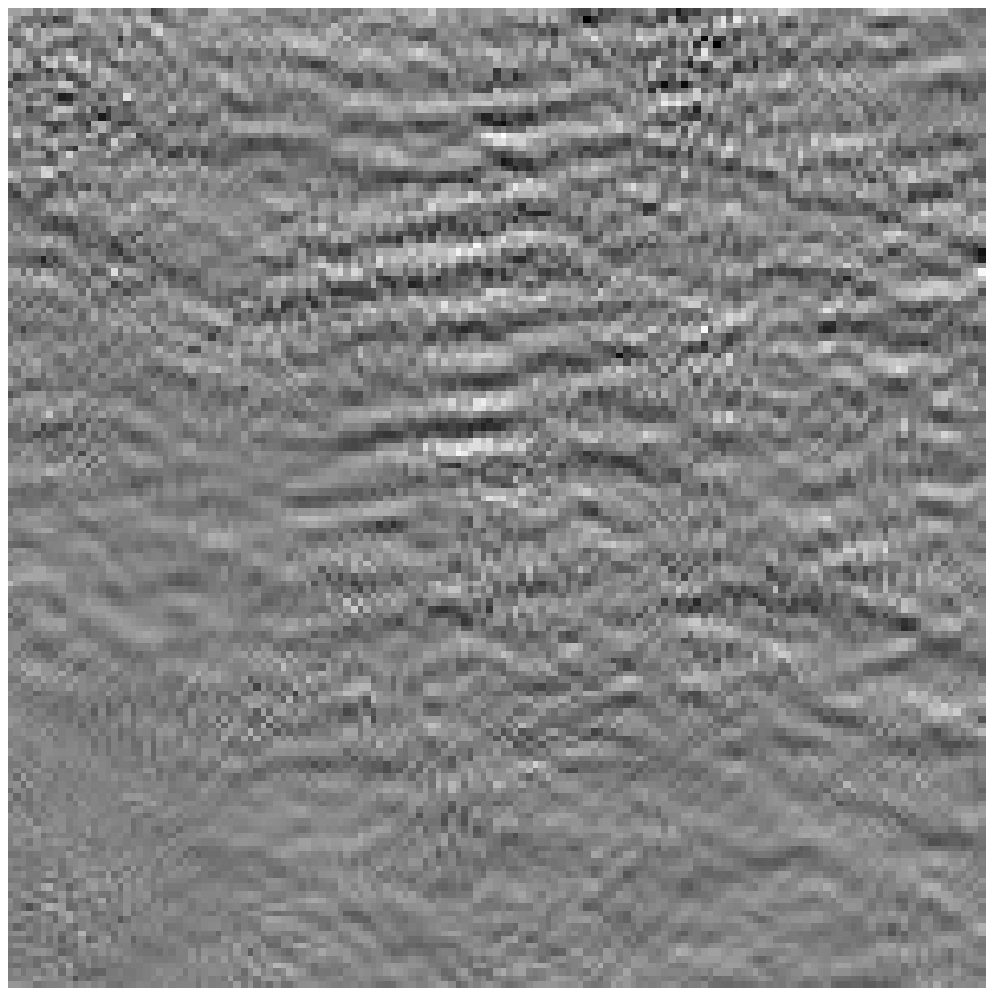}&
\includegraphics[width=8cm,height=4cm]{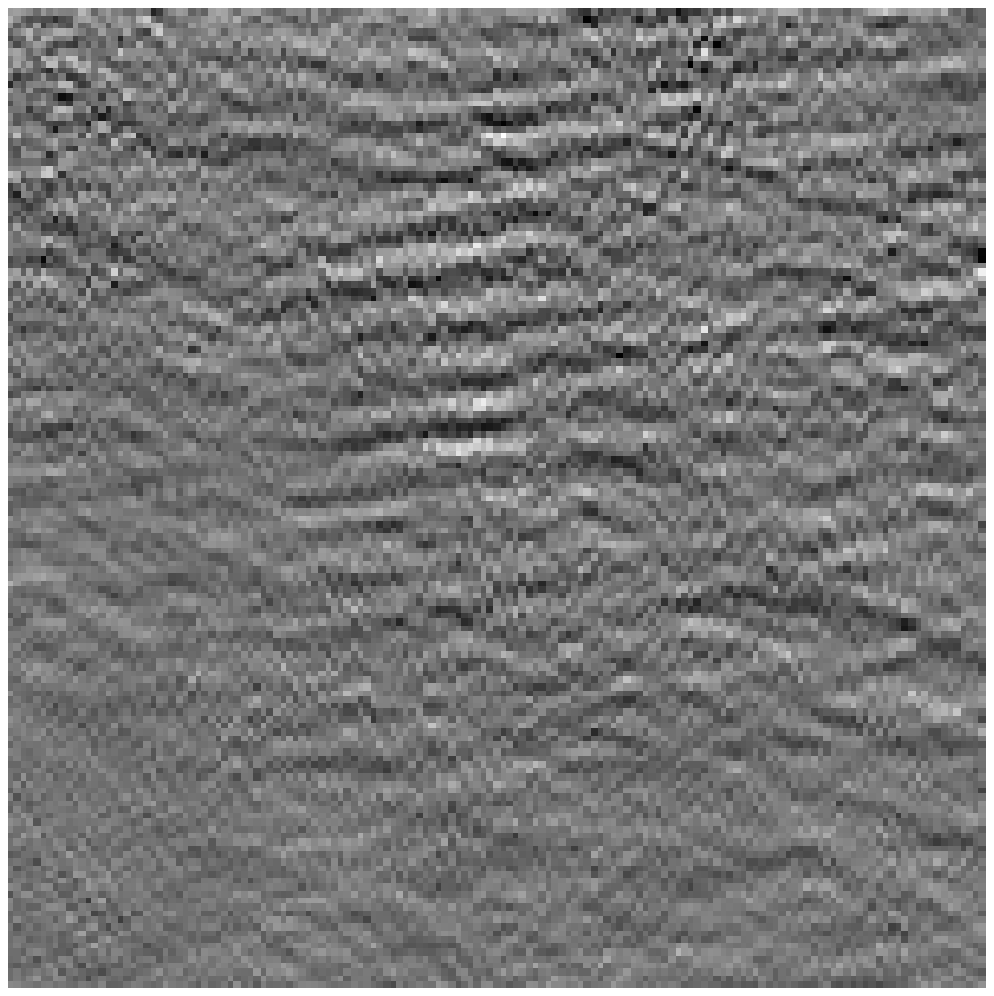}\\
(c) & (d) \\
\end{tabular}
\end{center}
\caption{Seismic data and denoising results  using Neighblock: (a) Original data; (b) Noisy data;  (c) DTT $M=2$; (d) DTT $M=4$. {\label{fig:agc512}}}
\end{figure}

\begin{itemize}
\item 
We have first applied our method on a $512\times 512$ directional texture image
(Straw D15 image from the Brodatz album) corrupted by an additive zero-mean white Gaussian noise.

The obtained SNR's (in dB) for three different initial noise levels are listed in Tab.~\ref{tab:simul_textu}.
We observe for this image that, by increasing the number of bands $M$, the denoising results are improved in almost all cases  for the DWT (sometimes only marginally) and significantly in almost all cases for the DTT.  
Furthermore, the DTT clearly leads to an improvement of the denoising performance compared with the DWT, whatever the initial SNR or the threshold selection method is. We remark that the more dramatic improvement over DWT is observed for Visushrink, which does not perform very well compared with SURE, NeighBlock or Bivariate.
Results are also relatively consistent between the top (noise variance known) and the bottom of the table (noise variance unknown), which is important in real applications where noise statistics often have to be estimated from the data.

Fig. \ref{fig:zoomtextu} also illustrates that, compared with other decompositions, the DTT with $M=4$ leads to sharper visual results and reduced artifacts. It can be seen from the bottom left corner that a 4-band DTT (Fig. \ref{fig:zoomtextu}d) better preserves the thin lines that are often blurred or merged in the other cropped images.

\item Second, we have performed the same denoising tests on the $512\times 512$ 8-bit Barbara image.
The obtained SNR's (in dB) are listed in Tab.~\ref{tab:simulsbarb}. 

For this image, we observe that, by increasing the number of bands $M$, the denoising results 
are improved in almost all cases both for the DWT and the DTT.  
Furthermore, the DTT clearly outperforms the DWT, as in the textured image case.

Fig. \ref{fig:zoombarb1} represents a zoom on a leg with a regular texture.  This illustrates that, compared with other decompositions, the 4-band DTT leads to better visual results.
Fig. \ref{fig:zoombarb1}a corresponding to the
2-band DWT
is strongly blurred. Details are better preserved in the 4-band decomposition (Fig. \ref{fig:zoombarb1}b), but it clearly appears that the texture with an apparent angle of  $\pi/4$ is heavily corrupted by patterns in the opposite direction, due to the mixing in the ``diagonal'' subband. Although Fig. \ref{fig:zoombarb1}c
remains blurry, there is much less directional mixture in both DTT decompositions.

\item Finally, we have tested our method on a $512\times 512$ seismic image displayed in Fig. \ref{fig:agc512}a. The data exhibits mostly horizontal structures as well as other directions which are important to the geophysicist for the underground analysis.

Similarly to previous cases, the seismic image is corrupted by an additive white Gaussian noise. The obtained denoising results  are listed in Tab.~\ref{tab:simulsagc512}.

 We observe that in most of the cases, denoising improves objectively   with the increase of the number of bands $M$, with DWT and DTT as well. Again, the best results are obtained with both dual-tree and a 4-band wavelet, but the gain over traditional DWT is sometimes smaller than in the previous example, for instance for NeighBlock shrinkage. It should be noted that the original image is not noise-free in general.  SNR measures are therefore more difficult to interpret. The existence of prior noise may explain the relatively weaker SNR increase between DWT and DTT, since  denoising  may attempt to remove both the added  and the original noise, and thus the denoised image strays away from the original noisy data.

 Fig. \ref{fig:agc512}b represents the original data corrupted with a -2 dB additive noise. Figs.  \ref{fig:agc512}c-d display the results with  2- and 4-band DTT respectively. 
Some of the oblique features (e.g. on the top-right corner) that are almost hidden in the noisy image become apparent in both the 2- and the 4-band DTT. 
We observe for this image that denoising results are more satisfactory with a 4-band than with  a 2-band DTT: the 2-band denoising image possesses larger blurred areas, especially in weakly energetic   zones. Careful examination also indicates a reduced presence of mosquito effects in the 4-band case.
\end{itemize}

We have experimented the DTT denoising algorithm on other image sets. Dual-tree $M$-band structures with $M> 2$ generally outperform existing wavelet
decompositions in terms of SNR. We shall remark that visual improvement is not always perceptible in image areas with weak directionality.

\subsection{Basis choice}

The previous section focused on the comparison between DWT and DTT with $M$-band Meyer wavelets, for different images, noise levels and threshold selection methods. Choosing a single wavelet family allowed us to provide a relatively fair comparison concerning the choice of the different aforementioned characteristics but it also appears interesting
to evaluate the influence of the decomposition filters. Amongst a variety of choices, we have tested $2$-band symmlets (with length 8), the basic 4-band Modulated Lapped Transform (MLT, see \cite{Malvar_H_1992_book_signal_plt}) and finally, Alkin and Caglar $4$-band filter bank \cite{Alkin_O_1995_j-ieee-tsp_design_embclpprp}. The results
concerning Meyer's wavelets can be found in previous tables.

The results reported in Tab.~\ref{tab:compare_ond} show 
the superiority of the $M$-band DTT (with $M>1$) over $M$-band
DWT or 2-band DTT, in particular when the popular symmlets are employed.
There is however no family which always leads to the best results.
We remark indeed that
DT MLT or AC DTT may lead to slightly improved results compared with
Meyer DTT,  but  the best choice often depends on the image.

\section{Conclusion}
\label{sec:conclu}
Motivated by applications where directional selectivity is of main interest, we have  proposed an extension of existing works 
on Hilbert transform pairs of dyadic orthonormal wavelets  to the 
$M$-band case.
In this context, we have pointed out that,
when several wavelet decompositions are performed in parallel,
special care should be taken concerning their implementation,
by designing appropriate pre- and post-processing stages.
Since the decomposition is redundant, an optimal reconstruction
has also been proposed.

By taking advantage of the Hilbert pair conditions and $M$-band features which offer
 additional degrees of freedom, 
this new transform has been applied to image denoising. 
Various simulations have allowed us to conclude that 
dual-tree decompositions with more than two bands generally outperform
discrete orthonormal wavelet decompositions and dyadic dual-tree
representations.

Encouraged by these results, we will consider further improvements with other filter bank designs, including regularity, as well as applications of  dual-tree $M$-band wavelets to other signal and image processing tasks, especially in seismics.
\begin{appendices}

\section{Proof of Proposition \ref{p:phase}:}
\label{ap:phase}
 Assuming that $\tilde{\theta}_0$ verifies the linearity relation \eqref{eq:linear}
 and using the fact that it is an odd function, we find that
\begin{multline}
\forall\; \omega \in ]-2\pi,2\pi[, \\
\beta(\omega) = \sum_{i=1}^\infty
\tilde{\theta}_0(\frac{\omega}{M^i})=\gamma\frac{\omega}{M}\sum_{i=0}^\infty\frac{1}{M^i}=\frac{\gamma\omega}{M-1}.
\label{eq:expbetalin}
\end{multline}
We deduce from Eq. \eqref{eq:SelesM} that, for all $m\in \{1,\ldots,M-1\}$,
\begin{multline}
\forall\; \omega \in ]-\frac{2\pi}{M},\frac{2\pi}{M}[ \\
\tilde{\alpha}_{0,m}(\omega)=\frac{\pi}{2}\textrm{sign}(\omega)-\frac{\gamma\omega
M}{M-1}\pmod{2\pi}.
\end{multline}
Furthermore, according to Condition $(\text{\textsf{P}}_{0,m})$,
\begin{equation}
\forall\; \omega \in ]-\frac{2\pi}{M},0] \qquad
\tilde{\alpha}_{0,m}(\omega+\frac{2\pi}{M})=\tilde{\alpha}_{0,m}(\omega)\pmod{2\pi}.
\end{equation}
This allows us to claim that there exists $d \in \ZZ$
such that
\begin{equation}
\gamma= (d+\frac{1}{2})(M-1).
\end{equation}
This leads to the expression of $\tilde{\alpha}_{0,m}$ in Eq. \eqref{eq:alpha0mlin}.
As $\tilde{\alpha}_{0,m}$ is a $2\pi/M$-periodic function, it is
 fully defined by its expression on $[0,\frac{2\pi}{M}[$. In contrast, we
 have to determine the expression of $\tilde{\theta}_0$ outside the interval
 $]-\frac{2\pi}{M},\frac{2\pi}{M}[$.
Using Eqs \eqref{eq:SelesM} and \eqref{eq:addscale}, we obtain, for all $m\in \{0,\ldots,M-1\}$,
\begin{align}
& \tilde{\alpha}_{0,m}(\frac{\omega}{M})+\tilde{\theta}_0(\frac{\omega}{M})+\beta(\frac{\omega}{M})=\frac{\pi}{2}\textrm{sign}(\omega) \pmod{2\pi}\nonumber\\
\Longleftrightarrow\; &\tilde{\theta}_0(\omega)=\frac{\pi}{2}\textrm{sign}(\omega)-\beta(\omega)-\tilde{\alpha}_{0,m}(\omega) \pmod{2\pi}.
\label{eq:theta0_beta}
\end{align}
Consider now the interval $[p\frac{2\pi}{M},(p+1)\frac{2\pi}{M}[$ where $p
 \in \Big\{1,\ldots,\Big\lceil  \frac{M}{2}\Big\rceil-1\Big\}$. 
As $[p\frac{2\pi}{M},(p+1)\frac{2\pi}{M}[ \subset [0,2\pi[ $,
Eq.~\eqref{eq:expbetalin} yields
\begin{equation}
\forall \omega \in [p\frac{2\pi}{M},(p+1)\frac{2\pi}{M}[,\qquad
\beta(\omega)=(d+\frac{1}{2})\omega.
\end{equation}
Using Eq.~\eqref{eq:theta0_beta} and
the $2\pi/M$-periodicity of $\tilde{\alpha}_{0,m}$, we deduce that
\begin{multline}
\forall \omega \in [p\frac{2\pi}{M},(p+1)\frac{2\pi}{M}[, \\
\tilde{\theta}_0(\omega)
=\frac{\pi}{2}-(d+\frac{1}{2})\omega-\tilde{\alpha}_{0,m}(\omega-\frac{2\pi}{M}p)
\pmod{2\pi}.
\end{multline}
Combining this result with Eq.~\eqref{eq:alpha0mlin} leads to Eq. \eqref{eq:theta0}. As a consequence of the antisymmetry of the phase of a real filter, a similar expression is obtained for $p \in \Big\{\Big\lceil\frac{M}{2}\Big\rceil,\ldots,M-1\Big\}$:
\begin{multline}
\forall \omega \in \Big]p \frac{2\pi}{M},(p+1)\frac{2\pi}{M}\Big], \\
\tilde{\theta}_0(\omega) = (d+\frac {1}{2})(M-1)\,\omega - p\pi
\pmod{2\pi}.
\label{eq:theta0b}
\end{multline}

In summary, under the considered assumptions, we have seen that, if there exists a solution to Eq.~\eqref{eq:SelesM},
it is given by Eqs.~\eqref{eq:alpha0mlin} and \eqref{eq:theta0}.
Conversely, we will now prove that any filters satisfying
Eqs.~\eqref{eq:alpha0mlin} and \eqref{eq:theta0}
are solutions to Eq. \eqref{eq:SelesM}.
More precisely, we will proceed by induction to show that
\begin{align}
\forall k \in \NN,\; &\forall\omega\in]2k\pi,2(k+1)\pi[, \nonumber \\ 
&\beta(\omega)=(d+\frac{1}{2})\omega-k\pi\pmod{2\pi}
\label{eq:inductbeta}\\
\mbox{and}\;\;
&\tilde{\alpha}_{0,m}(\frac{\omega}{M})+\beta(\omega)=\frac{\pi}{2} \pmod{2\pi}.
\label{eq:inductalpha}
\end{align}
\begin{itemize}
\item It is readily checked that the properties 
\eqref{eq:inductbeta}-\eqref{eq:inductalpha} are satisfied for $k=0$.
\item Assuming that the properties hold true up to the index $k-1\geq 0$, we will
demonstrate it remains valid at index $k$.\\
\noindent We can write $k=Mp+q$ with $p\in \mathbb{N}$ and $q\in \{0,...,M-1\}$
and, consequently,
\begin{align}
&\omega \in]2k\pi,2(k+1)\pi[\, \Longleftrightarrow \\
&\frac{\omega}{M} \in
]2(p+\frac{q}{M})\pi,2(p+\frac{q+1}{M})\pi[\,\subset\,]2p\pi,2(p+1)\pi[.\nonumber
\end{align}
Since $p<k$, according to the induction hypothesis, we have $\forall \omega \in]2k\pi,2(k+1)\pi[$,
\begin{equation}
\beta(\frac{\omega}{M})=(d+\frac{1}{2})\frac{\omega}{M}-p\pi \pmod{2\pi}.
\label{eq:expbetaomegaM} 
\end{equation}
Moreover, the $2\pi$-periodicity of $\tilde{\theta}_0$ allows us to write:
\begin{equation}
\\\tilde{\theta}_0(\frac{\omega}{M})=\\\tilde{\theta}_0(\frac{\omega}{M}-2p\pi).
\end{equation}
As $\omega/M-2p\pi \,\in\, ]2q\frac{\pi}{M},2(q+1)\frac{\pi}{M}[$,
Eqs.~\eqref{eq:theta0} and \eqref{eq:theta0b} lead to
\begin{align}
\tilde{\theta}_0(\frac{\omega}{M})&=\frac{M-1}{M}(d+\frac {1}{2})\omega \label{eq:exptheta0omageM}\\
&- \big((2d+1)(M-1)p+q\big)\pi \pmod{2\pi}\nonumber\\
& = \frac{M-1}{M}(d+\frac {1}{2})\omega-
(k-p)\pi \pmod{2\pi}.\nonumber
\end{align}
Combining Eqs.~\eqref{eq:addscale}, \eqref{eq:expbetaomegaM}
and  \eqref{eq:exptheta0omageM}, Eq.~\eqref{eq:inductbeta} is obtained.
By invoking the $2\pi/M$-periodicity of $\tilde{\alpha}_{0,m}$, the second part
of the property is proved in the similar way. Indeed, for $\omega \in]2k\pi,2(k+1)\pi[$, we have:
\begin{equation}
\tilde{\alpha}_{0,m}(\frac{\omega}{M})=\tilde{\alpha}_{0,m}(\frac{\omega}{M}-2(p+\frac{q}{M})\pi)
\end{equation}
which, using Eq.~\eqref{eq:alpha0mlin}, leads to
\begin{align}
\tilde{\alpha}_{0,m}(\frac{\omega}{M})&=\frac{\pi}{2}-(d+\frac{1}{2})M(\frac{\omega}{M}-2(p+\frac{q}{M})\pi) \nonumber \\
&=\frac{\pi}{2}-(d+\frac{1}{2})\omega+k\pi\pmod{2\pi}.
\end{align}
Then, summing Eq.~\eqref{eq:inductbeta} and the above expression results in Eq.~\eqref{eq:inductalpha}.
\end{itemize}
In conclusion, we have proved by induction that Eq.~\eqref{eq:inductalpha} holds for
almost all $\omega > 0$.
The function $\tilde{\theta}_0$ (and thus $\beta$) being odd as well as $\tilde{\alpha}_{0,m}$, we deduce that  Eq.~\eqref{eq:SelesM} is satisfied almost everywhere.
This ends the proof of Proposition \ref{p:phase}.

\section{Proof of Proposition \ref{prop:sym}:}
\label{ap:sym}
Assuming $h_0$ is symmetric w.r.t. $k_0$, we have
\begin{align}
&\forall k \in \ZZ,\qquad h_0[2k_0-k] = h_0[k]\\
\Longleftrightarrow\; & e^{-2\imath k_0 \omega} H_0^*(\omega)=H_0(\omega).
\end{align}
Thanks to Eq.~\eqref{eq:allpass}, this may be rewritten as
\begin{equation}
 e^{-2\imath k_0 \omega}e^{-2\imath \theta_0(\omega)}G_0^*(\omega)=G_0(\omega).
\end{equation}
According to Eq. \eqref{eq:theta0},
\begin{equation}
2\theta_0(\omega)= (2d+1)(M-1)\omega \pmod{2\pi}.
\end{equation}
which leads to
\begin{equation}
\forall k \in \ZZ,\qquad g_0[2k_0+(2d+1)(M-1)-k]  = g_0[k].
\end{equation}
This shows that
$g_0$ is symmetric w.r.t. $k_0+(d+\frac{1}{2})(M-1)$.

In the same way, for any $m \in \{1,...,M-1\}$, the symmetry/antisymmetry property:
\begin{equation}
\forall k \in \ZZ,\qquad 
h_m[2k_m-k]=\pm h_m[k]
\end{equation}
combined with Eq.~\eqref{eq:thetam}, results in:
\begin{equation}
\forall k \in \ZZ,\qquad
g_m[2k_m-2d-1-k] = \mp g_m[k].
\end{equation}

\section{Proof of Proposition \ref{prop:fram}:}
\label{ap:fram}
We denote by $\|.\|$ the norms of the underlying Hilbert spaces.
We have then, for all $\mathbf{f} \in  \ell^2(\ZZ^2)$,
\begin{equation}
\| \mathbf{D}\mathbf{f}\|^2 = \| \mathbf{D_1} \mathbf{f}\|^2
+ \| \mathbf{D_2} \mathbf{f}\|^2.
\end{equation}
Let us next focus on the first term on the right-hand side of this equation.
As $\mathbf{U_1}$ is unitary, we have
\begin{align}
\| \mathbf{D_1} \mathbf{f} \|^2 &= \| \mathbf{F_1} \mathbf{f} \|^2\label{eq:D1f2} \\
&= \frac{1}{(2\pi)^2} \int_{-\pi}^\pi\int_{-\pi}^\pi |F_1(\omega_x,\omega_y) 
\widehat{f}(\omega_x,\omega_y)|^2\,d\omega_x\,d\omega_y. \nonumber
\end{align}
In Equation~\eqref{eq:expF1}, we upper bound the magnitude of the sums by the sum of magnitudes. Invoking the Cauchy-Schwarz inequality, 
 the modulus of  the frequency response of
the first prefilter satisfies the following inequality:
\begin{multline}
|F_1(\omega_x,\omega_y)| 
\leq  \Bigl(\sum_{p,q}
|\widehat{s}(\omega_x+2p\pi,\omega_y+2q\pi)|^2\Bigr)^{1/2} \\ 
\hspace{-1cm}\Bigl(\sum_{p}|\widehat{\psi}_0(\omega_x+2p\pi)|^2\Bigr)^{1/2}
\Bigl(\sum_{q} |\widehat{\psi}_0(\omega_y+2q\pi)|^2\Bigr)^{1/2}.
\end{multline}
As $\{\psi_0(t-k),k\in \ZZ\}$ is an orthonormal family of $\LL^2(\RR)$,
$\sum_{p=-\infty}^\infty |\widehat{\psi}_0(\omega_x+2p\pi)|^2 = 1$.
Under the Assumptions~\eqref{eq:bounds} and \eqref{eq:boundpsi0}, we deduce that
\begin{equation}
|F_1(\omega_x,\omega_y)| \leq \sqrt{B_s^2+C_s}. 
\label{eq:MajF1}
\end{equation}
Besides, the frequency magnitude of the first prefilter can be lower bounded as
follows:
\begin{multline}
|F_1(\omega_x,\omega_y)| \geq |\widehat{s}(\omega_x,\omega_y)\widehat{\psi}_0(\omega_x)\widehat{\psi}_0(\omega_y)|\\ -\mathop{\sum_{(p,q)}}_{\neq (0,0)}
|\widehat{s}(\omega_x+2p\pi,\omega_y+2q\pi)\widehat{\psi}_0(\omega_x+2p\pi)
\widehat{\psi}_0(\omega_y+2q\pi)|.
\end{multline}
The latter summation can be upper bounded as we did for 
$|F_1(\omega_x,\omega_y)|$, which
combined with the assumptions~\eqref{eq:bounds} and
\eqref{eq:boundpsi0}, yields:
\begin{equation}
|F_1(\omega_x,\omega_y)| \geq
A_sA_{\psi_0}^2-\sqrt{C_s}.
\label{eq:MinF1}
\end{equation}
From Eqs. \eqref{eq:D1f2}, \eqref{eq:MajF1} and \eqref{eq:MinF1},
we conclude that
\begin{equation}
(A_s A_{\psi_0}^2-\sqrt{C_s}) \|\mathbf{f}\| \leq \| \mathbf{D_1} \mathbf{f} \| 
\leq  \sqrt{B_s^2+C_s}\|\mathbf{f}\|.
\label{eq:boundD1}
\end{equation}
Now, using Eq. \eqref{eq:expF2F1}
and invoking the same arguments as previously lead to
\begin{equation}
(A_s A_{\psi_0}^2-\sqrt{C_s}) \|\mathbf{f}\| \leq \| \mathbf{D_2}\mathbf{f} \| 
\leq \sqrt{B_s^2+C_s} \|\mathbf{f}\|.
\label{eq:boundD2}
\end{equation}
Combining Eqs.~\eqref{eq:boundD1} and \eqref{eq:boundD2} allows us to
conclude that
\begin{equation}
\sqrt{2}(A_s A_{\psi_0}^2-\sqrt{C_s}) \|\mathbf{f}\|\leq \| \mathbf{D}\mathbf{f}\| \leq
\sqrt{2(B_s^2+C_s)}\|\mathbf{f}\|.
\end{equation}
As we have assumed in Eq. \eqref{eq:boundpsi0} that $A_s A_{\psi_0}^2-\sqrt{C_s} > 0$, this means that $\mathbf{D}$ is a frame operator.
Note that, when ideal low-pass filters are used for $s$ and $\psi_0$
(that is $s(x,y) = \psi_0(x)\psi_0(y)$ with $\psi_0(t) = \mathrm{sinc}(\pi t)$),
we have $|F_1(\omega_x,\omega_y)|  = | F_2(\omega_x,\omega_y)| = 1$,
and thus, $\| \mathbf{D_1}\mathbf{f} \| =  \| \mathbf{D_2}\mathbf{f} \|
= \| \mathbf{f} \|$. Therefore, in this ideal case, $\mathbf{D}$ is a tight
frame operator with bound $\sqrt{2}$.

To determine the ``dual'' frame reconstruction operator, we have
to calculate the pseudo-inverse of $\mathbf{D}$ which is defined
by $\mathbf{D}^{\sharp} = (\mathbf{D}^{\dagger}\mathbf{D})^{-1}\mathbf{D}^{\dagger}$. 
In our case, the adjoint of $\mathbf{D}$ is
\begin{equation}
\mathbf{D}^{\dagger}=
(\mathbf{D_1}^{\dagger}\quad \mathbf{D_2}^{\dagger})=(\mathbf{F_1}^{\dagger}\mathbf{U_1}^{\dagger} \quad \mathbf{F_2}^{\dagger}\mathbf{U_2}^{\dagger}).
\end{equation}
Hence, by virtue of the unitarity of $\mathbf{U_1}$ and $\mathbf{U_2}$, we obtain
$\mathbf{D}^{\dagger}\mathbf{D}=\mathbf{F_1}^{\dagger}\mathbf{F_1}+
\mathbf{F_2}^{\dagger}\mathbf{F_2}$\\
and, finally,
\begin{equation}
\mathbf{D}^{\sharp} = 
(\mathbf{F_1}^{\dagger}\mathbf{F_1}+\mathbf{F_2}^{\dagger}\mathbf{F_2})^{-1} \; (\mathbf{F_1}^{\dagger}\mathbf{U_1}^{-1} \quad \mathbf{F_2}^{\dagger}
\mathbf{U_2}^{-1}).
\end{equation}
\end{appendices}

\vspace*{-1cm}
\bibliographystyle{IEEEtran}
%%\bibliography{abbr,gen200404,SP_General_New,Bib_MLV_New}
%\bibliography{abbr,Reference_Base_JabRef}

\onecolumn
\pagebreak
\thispagestyle{empty}
\listoffigures
\listoftables
\addtocounter{page}{-1}

\newpage
%%%%%%%%%%%%%%%%%%%%%%%%%%%%%%%%%%%%%%%%%%%%%%%%%%%%%%%%%%%%%%%%%%%%%%%%%%%%%%%%%%%%%%%%%%%%%%%%%%%%%%%%
%figures and tables
%%%%%%%%%%%%%%%%%%%%%%%%%%%%%%%%%%%%%%%%%%%%%%%%%%%%%%%%%%%%%%%%%%%%%%%%%%%%%%%%%%%%%%%%%%%%%%%%%%%%%%%%

\begin{figure} 
\centering
\input{figures/Mbanddualtree.pstex_t} 
\caption{$M$-band dual-tree decomposition scheme over 2 resolution levels.}
\label{fig:Mbanddualtree}
\end{figure}
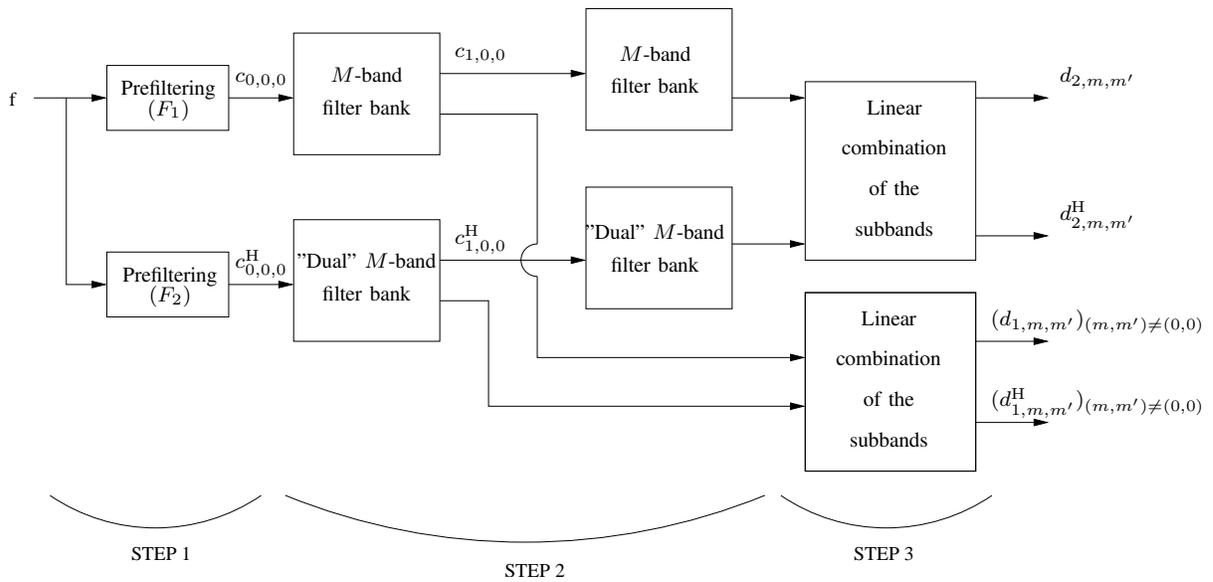

\input{simul_textu.tex}
\input{simuls4barb.tex}

\input{simuls4agc512.tex}

\input{simuls_compare_ond.tex}

\end{document}

%% file: figures/fb_M_ortho_dual.pstex_t
\begin{picture}(0,0)%
\includegraphics{fb_M_ortho_dual.pstex}%
\end{picture}%
\setlength{\unitlength}{3066sp}%
\begingroup\makeatletter\ifx\SetFigFont\undefined%
\gdef\SetFigFont#1#2#3#4#5{%
  \reset@font\fontsize{#1}{#2pt}%
  \fontfamily{#3}\fontseries{#4}\fontshape{#5}%
  \selectfont}%
\fi\endgroup%
\begin{picture}(5084,3419)(609,-8858)
\put(2026,-7441){\makebox(0,0)[b]{\smash{{\SetFigFont{9}{10.8}{\familydefault}{\mddefault}{\updefault}$G^*_0$}}}}
\put(2026,-7981){\makebox(0,0)[b]{\smash{{\SetFigFont{9}{10.8}{\familydefault}{\mddefault}{\updefault}$G^*_1$}}}}
\put(2026,-8701){\makebox(0,0)[b]{\smash{{\SetFigFont{9}{10.8}{\familydefault}{\mddefault}{\updefault}$G^*_{M-1}$}}}}
\put(2836,-7441){\makebox(0,0)[b]{\smash{{\SetFigFont{9}{10.8}{\familydefault}{\mddefault}{\updefault}$\downarrow\!M$}}}}
\put(2836,-7981){\makebox(0,0)[b]{\smash{{\SetFigFont{9}{10.8}{\familydefault}{\mddefault}{\updefault}$\downarrow\!M$}}}}
\put(2836,-8701){\makebox(0,0)[b]{\smash{{\SetFigFont{9}{10.8}{\familydefault}{\mddefault}{\updefault}$\downarrow\!M$}}}}
\put(4006,-7441){\makebox(0,0)[b]{\smash{{\SetFigFont{9}{10.8}{\familydefault}{\mddefault}{\updefault}$\uparrow\!M$}}}}
\put(4006,-7981){\makebox(0,0)[b]{\smash{{\SetFigFont{9}{10.8}{\familydefault}{\mddefault}{\updefault}$\uparrow\!M$}}}}
\put(4006,-8701){\makebox(0,0)[b]{\smash{{\SetFigFont{9}{10.8}{\familydefault}{\mddefault}{\updefault}$\uparrow\!M$}}}}
\put(4816,-7441){\makebox(0,0)[b]{\smash{{\SetFigFont{9}{10.8}{\familydefault}{\mddefault}{\updefault}$G_0$}}}}
\put(4816,-7981){\makebox(0,0)[b]{\smash{{\SetFigFont{9}{10.8}{\familydefault}{\mddefault}{\updefault}$G_1$}}}}
\put(4816,-8701){\makebox(0,0)[b]{\smash{{\SetFigFont{9}{10.8}{\familydefault}{\mddefault}{\updefault}$G_{M-1}$}}}}
\put(1801,-5686){\makebox(0,0)[b]{\smash{{\SetFigFont{9}{10.8}{\familydefault}{\mddefault}{\updefault}$H^*_0$}}}}
\put(1801,-6226){\makebox(0,0)[b]{\smash{{\SetFigFont{9}{10.8}{\familydefault}{\mddefault}{\updefault}$H^*_1$}}}}
\put(1801,-6946){\makebox(0,0)[b]{\smash{{\SetFigFont{9}{10.8}{\familydefault}{\mddefault}{\updefault}$H^*_{M-1}$}}}}
\put(2611,-5686){\makebox(0,0)[b]{\smash{{\SetFigFont{9}{10.8}{\familydefault}{\mddefault}{\updefault}$\downarrow\!M$}}}}
\put(2611,-6226){\makebox(0,0)[b]{\smash{{\SetFigFont{9}{10.8}{\familydefault}{\mddefault}{\updefault}$\downarrow\!M$}}}}
\put(3781,-5686){\makebox(0,0)[b]{\smash{{\SetFigFont{9}{10.8}{\familydefault}{\mddefault}{\updefault}$\uparrow\!M$}}}}
\put(3781,-6226){\makebox(0,0)[b]{\smash{{\SetFigFont{9}{10.8}{\familydefault}{\mddefault}{\updefault}$\uparrow\!M$}}}}
\put(3781,-6946){\makebox(0,0)[b]{\smash{{\SetFigFont{9}{10.8}{\familydefault}{\mddefault}{\updefault}$\uparrow\!M$}}}}
\put(4591,-5686){\makebox(0,0)[b]{\smash{{\SetFigFont{9}{10.8}{\familydefault}{\mddefault}{\updefault}$H_0$}}}}
\put(4591,-6226){\makebox(0,0)[b]{\smash{{\SetFigFont{9}{10.8}{\familydefault}{\mddefault}{\updefault}$H_1$}}}}
\put(4591,-6946){\makebox(0,0)[b]{\smash{{\SetFigFont{9}{10.8}{\familydefault}{\mddefault}{\updefault}$H_{M-1}$}}}}
\put(2611,-6946){\makebox(0,0)[b]{\smash{{\SetFigFont{9}{10.8}{\familydefault}{\mddefault}{\updefault}$\downarrow\!M$}}}}
\end{picture}%

%% file: figures/Mbanddualtree.pstex_t
\begin{picture}(0,0)%
\includegraphics{Mbanddualtree.pstex}%
\end{picture}%
\setlength{\unitlength}{2684sp}%
\begingroup\makeatletter\ifx\SetFigFont\undefined%
\gdef\SetFigFont#1#2#3#4#5{%
  \reset@font\fontsize{#1}{#2pt}%
  \fontfamily{#3}\fontseries{#4}\fontshape{#5}%
  \selectfont}%
\fi\endgroup%
\begin{picture}(10573,5262)(751,-7711)
\put(8926,-4561){\makebox(0,0)[b]{\smash{{\SetFigFont{8}{9.6}{\familydefault}{\mddefault}{\updefault}subbands}}}}
\put(8925,-4186){\makebox(0,0)[b]{\smash{{\SetFigFont{8}{9.6}{\familydefault}{\mddefault}{\updefault}of the}}}}
\put(8925,-3811){\makebox(0,0)[b]{\smash{{\SetFigFont{8}{9.6}{\familydefault}{\mddefault}{\updefault}combination}}}}
\put(8926,-3436){\makebox(0,0)[b]{\smash{{\SetFigFont{8}{9.6}{\familydefault}{\mddefault}{\updefault}Linear}}}}
\put(8874,-6511){\makebox(0,0)[b]{\smash{{\SetFigFont{8}{9.6}{\familydefault}{\mddefault}{\updefault}subbands}}}}
\put(8873,-6136){\makebox(0,0)[b]{\smash{{\SetFigFont{8}{9.6}{\familydefault}{\mddefault}{\updefault}of the}}}}
\put(8873,-5761){\makebox(0,0)[b]{\smash{{\SetFigFont{8}{9.6}{\familydefault}{\mddefault}{\updefault}combination}}}}
\put(8874,-5386){\makebox(0,0)[b]{\smash{{\SetFigFont{8}{9.6}{\familydefault}{\mddefault}{\updefault}Linear}}}}
\put(4050,-3136){\makebox(0,0)[b]{\smash{{\SetFigFont{8}{9.6}{\familydefault}{\mddefault}{\updefault}$M$-band}}}}
\put(4050,-3436){\makebox(0,0)[b]{\smash{{\SetFigFont{8}{9.6}{\familydefault}{\mddefault}{\updefault}filter bank}}}}
\put(4051,-4861){\makebox(0,0)[b]{\smash{{\SetFigFont{8}{9.6}{\familydefault}{\mddefault}{\updefault}"Dual" $M$-band}}}}
\put(4050,-5161){\makebox(0,0)[b]{\smash{{\SetFigFont{8}{9.6}{\familydefault}{\mddefault}{\updefault}filter bank}}}}
\put(6713,-3241){\makebox(0,0)[b]{\smash{{\SetFigFont{8}{9.6}{\familydefault}{\mddefault}{\updefault}filter bank}}}}
\put(6713,-2941){\makebox(0,0)[b]{\smash{{\SetFigFont{8}{9.6}{\familydefault}{\mddefault}{\updefault}$M$-band}}}}
\put(6713,-4891){\makebox(0,0)[b]{\smash{{\SetFigFont{8}{9.6}{\familydefault}{\mddefault}{\updefault}filter bank}}}}
\put(6713,-4591){\makebox(0,0)[b]{\smash{{\SetFigFont{8}{9.6}{\familydefault}{\mddefault}{\updefault}"Dual" $M$-band}}}}
\put(2219,-3444){\makebox(0,0)[b]{\smash{{\SetFigFont{8}{9.6}{\familydefault}{\mddefault}{\updefault}$(F_1)$}}}}
\put(2221,-3259){\makebox(0,0)[b]{\smash{{\SetFigFont{8}{9.6}{\familydefault}{\mddefault}{\updefault}Prefiltering}}}}
\put(2236,-4984){\makebox(0,0)[b]{\smash{{\SetFigFont{8}{9.6}{\familydefault}{\mddefault}{\updefault}Prefiltering}}}}
\put(2234,-5169){\makebox(0,0)[b]{\smash{{\SetFigFont{8}{9.6}{\familydefault}{\mddefault}{\updefault}$(F_2)$}}}}
\put(10801,-3136){\makebox(0,0)[b]{\smash{{\SetFigFont{8}{9.6}{\familydefault}{\mddefault}{\updefault}$d_{2,m,m'}$}}}}
\put(10801,-4411){\makebox(0,0)[b]{\smash{{\SetFigFont{8}{9.6}{\familydefault}{\mddefault}{\updefault}$d^{\mathrm{H}}_{2,m,m'}$}}}}
\put(10801,-5386){\makebox(0,0)[b]{\smash{{\SetFigFont{8}{9.6}{\familydefault}{\mddefault}{\updefault}$(d_{1,m,m'})_{(m,m')\neq(0,0)}$}}}}
\put(10801,-6136){\makebox(0,0)[b]{\smash{{\SetFigFont{8}{9.6}{\familydefault}{\mddefault}{\updefault}$(d^{\mathrm{H}}_{1,m,m'})_{(m,m')\neq(0,0)}$}}}}
\put(3076,-4861){\makebox(0,0)[b]{\smash{{\SetFigFont{8}{9.6}{\familydefault}{\mddefault}{\updefault}$c^{\mathrm{H}}_{0,0,0}$}}}}
\put(3076,-3136){\makebox(0,0)[b]{\smash{{\SetFigFont{8}{9.6}{\familydefault}{\mddefault}{\updefault}$c_{0,0,0}$}}}}
\put(5101,-2911){\makebox(0,0)[b]{\smash{{\SetFigFont{8}{9.6}{\familydefault}{\mddefault}{\updefault}$c_{1,0,0}$}}}}
\put(5101,-4636){\makebox(0,0)[b]{\smash{{\SetFigFont{8}{9.6}{\familydefault}{\mddefault}{\updefault}$c^{\mathrm{H}}_{1,0,0}$}}}}
\end{picture}%

%% file: figures/rot.pstex_t
\begin{picture}(0,0)%
\includegraphics{rot.pstex}%
\end{picture}%
\setlength{\unitlength}{2763sp}%
\begingroup\makeatletter\ifx\SetFigFont\undefined%
\gdef\SetFigFont#1#2#3#4#5{%
  \reset@font\fontsize{#1}{#2pt}%
  \fontfamily{#3}\fontseries{#4}\fontshape{#5}%
  \selectfont}%
\fi\endgroup%
\begin{picture}(3624,3693)(589,-3373)
\put(976,-1486){\makebox(0,0)[b]{\smash{{\SetFigFont{8}{9.6}{\familydefault}{\mddefault}{\updefault}$-\pi$}}}}
\put(2176,-2986){\makebox(0,0)[b]{\smash{{\SetFigFont{8}{9.6}{\familydefault}{\mddefault}{\updefault}$-\pi$}}}}
\put(3751,-1486){\makebox(0,0)[b]{\smash{{\SetFigFont{8}{9.6}{\familydefault}{\mddefault}{\updefault}$\pi$}}}}
\put(2251,-211){\makebox(0,0)[b]{\smash{{\SetFigFont{8}{9.6}{\familydefault}{\mddefault}{\updefault}$\pi$}}}}
\put(2176,164){\makebox(0,0)[b]{\smash{{\SetFigFont{8}{9.6}{\familydefault}{\mddefault}{\updefault}$\omega_y$}}}}
\put(4201,-1711){\makebox(0,0)[b]{\smash{{\SetFigFont{8}{9.6}{\familydefault}{\mddefault}{\updefault}$\omega_x$}}}}
\end{picture}%

%% file: simul_textu.tex
\begin{table}[htbp]
\begin{center}
\begin{tabular}{|c| c| c| c| c|| c| c| c| c|| c| c| c| c|} %| c| c| c| c|}
%barb
%sigma=25,30,35
%bruit estimé
\hline
 & \multicolumn{4}{|c||}{$\textrm{SNR}_{init} = 7.71$ dB} & \multicolumn{4}{|c||}{$\textrm{SNR}_{init} = 5.71$ dB} & \multicolumn{4}{|c|}{$\textrm{SNR}_{init} = 3.71$ dB}\\% & \multicolumn{4}{|c|}{$\textrm{SNR}_{init} = 2.67$ dB}\\
\hline
 & Visu & SURE & Biv & NB  & Visu & SURE  & Biv & NB & Visu & SURE & Biv & NB  \\% & Visu & SURE & Biv & NB  \\
\hline
DWT $M=2$ &5.44 &10.07 &10.37 &10.72  &4.36 &8.70 &9.02 &9.40  &3.37 &7.49 &7.75 &8.14  \\ %&3.12 &6.35 &6.57 &6.95  \\
\hline
DWT $M=3$ &5.57 &10.25 &10.38 &10.86  &4.53 &8.82 &9.01 &9.52  &3.62 &7.52 &7.72 &8.24  \\ %  &3.41 &6.41 &6.51 &7.03 \\
\hline
DWT $M=4$  &5.53 &10.25 &10.38 &10.94  &4.43 &8.83 &9.03 &9.59  &3.44 &7.65 &7.75 &8.31  \\ % &3.29 &6.45 &6.54 &7.08  \\
\hline
 DTT $M=2$ &6.67 &10.67 &10.85 &11.01  &5.51 &9.38 &9.54 &9.70  &4.39 &8.12 &8.29 &8.46  \\ % &4.05 &6.94  &7.10 &7.32\\
\hline
DTT $M=3$ &6.72 &10.80 &10.93 &11.19  &5.54 &9.47 &9.60 &9.85  &\textbf{4.54} &8.15 &8.33 &8.57  \\ %  &4.29 &7.02 &7.11 &7.34 \\
\hline
DTT $M=4$ &\textbf{6.91} &\textbf{10.91} &\textbf{10.96} &\textbf{11.31}  &\textbf{5.64} &\textbf{9.50} &\textbf{9.65} &\textbf{9.98}  &4.48 &\textbf{8.28} &\textbf{8.40} &\textbf{8.69} \\ %  &4.23&7.13 &7.45 &7.20\\
\hline
\hline
DWT $M=2$ &4.78 &9.71 & 9.99 &10.49  &3.94 &8.56 &8.78 &9.30  & 3.13 &7.41 &7.60 &8.12  \\ %  &2.98&6.10 &6.48 &6.97 \\
\hline
DWT $M=3$ &5.18 &9.96 &10.29 &10.80  &4.29 &8.59 &8.95 &9.51  &3.49 &7.50 &7.68 &8.26  \\ %  &3.34&6.40 &6.56 &7.06 \\
\hline
DWT $M=4$ &5.20 &10.04 &10.40 &10.90   &4.22 &8.78 &9.04 &9.59  &3.32 &7.63 &7.75 &8.32  \\ % &3.22 &6.33 &6.62 &7.11 \\
\hline
DTT $M=2$ &5.91 &10.33 &10.53 &10.86  &4.98 &9.15 &9.32 &9.66  &4.04 &8.04 & 8.14 &8.48  \\ %&4.01 &6.59 &7.08 &7.32  \\
\hline
DTT $M=3$ &6.23 &10.45 &10.87 &11.17  &5.25 &9.22 &9.56 &9.87  & \textbf{4.37} &8.06 &8.29 &8.60  \\ % &4.20 &7.02 &7.04 &7.38 \\
\hline
DTT $M=4$ &\textbf{6.52} &\textbf{10.62} &\textbf{10.99} &\textbf{11.31}  &\textbf{5.40} &\textbf{9.45} &\textbf{9.68} &\textbf{10.00}   &4.33 &\textbf{8.23} &\textbf{8.42} &\textbf{8.73} \\ % &4.15 &7.03 &7.21 &7.49  \\
\hline
\end{tabular}
\end{center}
\caption{Denoising results on texture image for different initial SNR's. In the top
part of the table, the variance is assumed to be known and in the
bottom one, it is estimated. The considered estimators are
SureShrink (SURE) \cite{Donoho_Johnstone1995}, Neighblock (NB)
\cite{Cai_Silverman2001}, Bivariate Shrinkage (Biv)
\cite{Sendur_Selesnick2002} and Visushrink (Visu).
\label{tab:simul_textu}}
\end{table}

%% file: simuls4barb.tex
\begin{table}[htbp]
\begin{center}
\begin{tabular}{|c| c| c| c| c|| c| c| c| c|| c| c| c| c|}
%barb
%sigma=25,30,35
%bruit estimé
\hline
 & \multicolumn{4}{|c||}{$\textrm{SNR}_{init} = 5.67$ dB} & \multicolumn{4}{|c||}{$\textrm{SNR}_{init} = 4.17$ dB} & \multicolumn{4}{|c|}{$\textrm{SNR}_{init} = 2.67$ dB}\\
\hline
 & Visu & SURE & Biv & NB  & Visu & SURE & Biv & NB  & Visu & SURE & Biv & NB   \\
\hline
%bruit connu
%im=sure
  DWT $M=2$  & 8.67 & 12.21 & 13.27 & 13.44  & 8.18 & 10.90 & 12.30 & 12.49   & 7.83 & 10.15 & 11.37 & 11.57  \\
  \hline
 DWT $M=3$  & 9.65 & 12.18 & 13.32 & 13.52  & 9.06 & 11.13 & 12.41& 12.59  & 8.53 & 10.43 & 11.54 & 11.68  \\
  \hline
 DWT $M=4$ & 9.65 & 12.60 & 13.37 & 13.65  & 9.01 & 11.03 & 12.51 & 12.73  & 8.42 & 10.39 & 11.68 & 11.83  \\
  \hline
 DTT $M=2$ & 9.38 & 12.89 & 13.76 & 13.69  & 8.73 & 11.93 & 12.79 & 12.74  & 8.25 & 10.88 & 11.84 & 11.80  \\
  \hline
  DTT $M=3$ & 10.45 & 12.80 & 13.99 & 13.83  & 9.66 & 11.69 & 13.06 & 12.88  & 8.97 & 10.95 & 12.15 & 11.93  \\
  \hline
 DTT $M=4$ &\textbf{ 10.80} &\textbf{13.32} & \textbf{14.16} &\textbf{ 14.01}  &\textbf{ 10.05} &\textbf{ 12.28} &\textbf{ 13.31} &\textbf{ 13.07}  &\textbf{ 9.35} &\textbf{ 11.20} &\textbf{ 12.47} &\textbf{ 12.15}  \\
 \hline
\hline
 DWT $M=2$ & 8.63 & 12.19 & 13.25 & 13.50   & 8.16 & 10.89  & 12.28 & 12.55 & 7.82 & 10.14 & 11.35 & 11.62  \\
 \hline
 DWT $M=3$ & 9.63 & 12.17 & 13.31 & 13.55   & 9.05 & 11.13 & 12.41 & 12.61  & 8.53 & 10.42 & 11.54 & 11.70  \\
  \hline
 DWT $M=4$ & 9.62 & 12.55 & 13.37 & 13.68   & 8.99 & 11.04 & 12.51 & 12.76  & 8.41 & 10.39 & 11.68 & 11.86 \\
  \hline
 DTT $M=2$  & 9.33 & 12.88 & 13.74 & 13.75  & 8.70 & 11.92 & 12.77 & 12.79   & 8.23 & 10.85 & 11.82 & 11.84 \\
 \hline
 DTT $M=3$ & 10.43 & 12.78 & 13.99 & 13.85  & 9.65 & 11.70 & 13.06 & 12.89  & 8.97 & 10.96 & 12.14 & 11.94  \\
 \hline
 DTT $M=4$ & \textbf{10.78} &\textbf{ 13.30} &\textbf{ 14.17} & \textbf{14.04}  &\textbf{ 10.04} & \textbf{12.23} &\textbf{ 13.31} &\textbf{ 13.10}   & \textbf{9.34} & \textbf{11.21} &\textbf{ 12.47} & \textbf{12.17}  \\
 \hline

\end{tabular}
\end{center}
\caption{Denoising results on Barbara image for different initial SNR's. In the top part
of the table, the variance is assumed to be known  and in the bottom one, it is
estimated. The considered estimators are SureShrink (SURE) \cite{Donoho_Johnstone1995}, Neighblock (NB) \cite{Cai_Silverman2001}, Bivariate Shrinkage (Biv) \cite{Sendur_Selesnick2002} and Visushrink (Visu). \label{tab:simulsbarb}}
\end{table}

%% file: simuls4agc512.tex
\begin{table}[htbp]
\begin{center}
\begin{tabular}{|c| c| c| c| c|| c| c| c| c|| c| c| c| c|}
%agc512
%snr=4,3,2
\hline
 & \multicolumn{4}{|c||}{$\textrm{SNR}_{init} = 4.13$ dB} & \multicolumn{4}{|c||}{$\textrm{SNR}_{init} = 3.13$ dB} & \multicolumn{4}{|c|}{$\textrm{SNR}_{init} = 2.13$ dB}\\
\hline
 & Visu & SURE & Biv & NB   & Visu & SURE & Biv & NB  & Visu & SURE & Biv & NB  \\
\hline
%bruit connu
DWT $M=2$  & 3.17 & 6.66 & 6.78 & 7.46  & 2.83 & 6.05  & 6.19 & 6.87  & 2.51 & 5.48 & 5.64 & 6.30 \\
\hline
DWT $M=3$ & 3.53 & 7.12 & 7.14 & 7.84  & 3.21  & 6.51 & 6.53 & 7.23  & 2.90 & 5.91 & 5.96 & 6.64  \\
\hline
 DWT $M=4$ & 3.60 & 7.52 & 7.47 & 8.16   & 3.24 & 6.91 & 6.83 & 7.53   & 2.91 & 6.31 & 6.23 & 6.93 \\
\hline
  DTT $M=2$ & 3.82 & 7.12 & 7.10 & 7.57  & 3.47 & 6.52 & 6.50 & 6.98  & 3.12 & 5.96 & 5.96 & 6.42  \\
\hline
DTT $M=3$ & 4.15 & 7.49 & 7.42 & 7.92  & 3.79 & 6.91 & 6.82 & 7.31   & 3.46 & 6.28 & 6.25 & 6.72  \\
\hline
DTT $M=4$ &\textbf{ 4.23} & \textbf{7.82} & \textbf{7.72} & \textbf{8.21}  & \textbf{3.84} & \textbf{7.23} & \textbf{7.09} &\textbf{ 7.58}   & \textbf{3.49} & \textbf{6.65} & \textbf{6.49} & \textbf{6.98}  \\
\hline
 \hline
DWT $M=2$ & 2.56 & 5.19 & 5.73 & 6.76  & 2.34 & 4.92 & 5.37 & 6.34  & 2.11 & 4.64 & 5.03 & 5.92  \\
\hline
DWT $M=3$ & 3.27 & 6.60  & 6.77 & 7.72  & 3.01 & 6.28 & 6.26 & 7.16  & 2.75 & 5.62 & 5.76 & 6.61  \\
\hline
 DWT $M=4$  & 3.50 & 7.51 & 7.36 & 8.16  & 3.17 & 6.88 & 6.74 & 7.54  & 2.86 & 6.29 & 6.15 & 6.94  \\
\hline
DTT $M=2$  & 3.12 & 5.86 & 5.97 & 6.93  & 2.89 & 5.51 & 5.62 & 6.51   & 2.65 & 4.95 & 5.28 & 6.10  \\
\hline
DTT $M=3$ & 3.84 & 7.07 & 7.04 & 7.84  & 3.55 & 6.56 & 6.52 & 7.27  & 3.27 & 5.97 & 6.02 & 6.72  \\
\hline
DTT $M=4$ &\textbf{ 4.11} & \textbf{7.81} & \textbf{7.60} & \textbf{8.23}  & \textbf{3.76} & \textbf{7.22} &\textbf{ 6.99} & \textbf{7.60}   & \textbf{3.42} & \textbf{6.64} & \textbf{6.41} & \textbf{7.00}  \\
\hline
 \end{tabular}
 \end{center}
\caption{Denoising results on seismic image for different initial SNR's. In the top part
of the table, the variance is assumed to be known  and in the bottom one, it is
estimated. The considered estimators are SureShrink (SURE) \cite{Donoho_Johnstone1995}, Neighblock (NB) \cite{Cai_Silverman2001}, Bivariate Shrinkage (Biv) \cite{Sendur_Selesnick2002}  and Visushrink (Visu). \label{tab:simulsagc512}}
 \end{table}

%% file: simuls_compare_ond.tex
\begin{table}
\begin{center}
\begin{tabular}{|c| c| c| c| c|| c| c| c| c|| c| c| c| c|}
\hline
  & Visu & SURE & Biv & NB & Visu & SURE & Biv & NB & Visu & SURE & Biv & NB \\
\hline
\textbf{Texture} & \multicolumn{4}{|c||}{$\textrm{SNR}_{init} = 7.71$ dB} &  \multicolumn{4}{|c||}{$\textrm{SNR}_{init} = 5.71$ dB} &  \multicolumn{4}{|c|}{$\textrm{SNR}_{init} = 3.71$ dB}\\
\hline
 symlet  DWT & 5.01 & 9.78 & 9.96 & 10.33  & 3.97 & 8.40 & 8.58 & 8.99  & 3.07 &  7.12 & 7.31 & 7.73 \\
\hline
 DW MLT  & 5.04 &  10.08 & 10.11 & 10.58  & 3.94 & 8.60 & 8.71 & 9.20  &  3.01 &  7.33 &  7.38 &  7.89  \\
\hline
 AC  DWT  & 5.18 &10.06 & 10.07 &10.58 & 4.11 & 8.61 &8.70 & 9.22  &3.19 &  7.32 & 7.39 & 7.94 \\
\hline
 symlet DTT &  6.59 &10.64  & 10.85 &10.91 & 5.36 & 9.36 & 9.55 & 9.61   & 4.24 & 8.16 & 8.32 & 8.38 \\
\hline
 DT MLT   & 6.94 & \textbf{ 11.04} &\textbf{ 11.07} & \textbf{11.32} &  5.56 & \textbf{ 9.72} & \textbf{ 9.79} & \textbf{ 9.99}  &  4.35 & \textbf{ 8.50} &  \textbf{ 8.54} & 8.70  \\
\hline
 AC DTT & \textbf{6.95} &10.97 & 11.01 &11.29  &\textbf{5.60}  &9.69 & 9.74 & 9.97  &\textbf{ 4.40}  & 8.45 & 8.52 &\textbf{ 8.71}   \\
\hline
\hline
\textbf{Barbara} & \multicolumn{4}{|c||}{$\textrm{SNR}_{init} = 5.67$ dB} & \multicolumn{4}{|c||}{$\textrm{SNR}_{init} = 4.17$ dB} & \multicolumn{4}{|c|}{$\textrm{SNR}_{init} = 2.67$ dB}\\
\hline
 symlet  DWT & 8.66 & 11.83 &12.72 &12.95  & 8.21 & 10.76 &11.83 & 12.06 & 7.85 & 9.94 &10.98 &11.19 \\
\hline
 DW MLT & 8.95 & 12.05 & 12.70 & 12.96  & 8.37 & 11.00 & 11.81 & 12.05   & 7.88 & 9.81 & 10.97 & 11.17 \\
\hline
 AC  DWT  &  9.20 & 12.17 &12.93 & 13.17  & 8.58 & 10.86 &12.06 & 12.27  & 8.08  & 9.94 &11.23 &11.39 \\
\hline
 symlet DTT  & 9.45 & 12.92 &13.69 &13.62 & 8.86& 11.82 & 12.74 & 12.70 &8.43 & 10.85 & 11.83 &11.80   \\
\hline
 DT MLT  & 10.49 & 13.29 & 14.15 & 13.98  & 9.67 &\textbf{ 12.32} & 13.26 & 13.07    & 8.94 & 11.07  & 12.39 & 12.17\\
\hline
 AC  DTT &\textbf{ 10.71} &\textbf{ 13.40} &\textbf{ 14.31} &\textbf{14.08}  &\textbf{9.88}  & 12.31 &\textbf{ 13.43} &\textbf{ 13.17} &\textbf{9.12}  &\textbf{ 11.16} &\textbf{12.56} &\textbf{ 12.28}  \\
 \hline
\hline
\textbf{Seismic} & \multicolumn{4}{|c||}{$\textrm{SNR}_{init} = 4.13$ dB} & \multicolumn{4}{|c||}{$\textrm{SNR}_{init} = 3.13$ dB} & \multicolumn{4}{|c|}{$\textrm{SNR}_{init} = 2.13$ dB}\\
\hline
 symlet  DWT & 3.22 & 6.64 & 6.74 & 7.39  & 2.91& 6.04 & 6.15 & 6.80    &2.60 & 5.47 &5.60 & 6.23  \\
\hline
 DW MLT &  3.54 & 7.09 &  7.08 &  7.72  &  3.22 &  7.11 &  6.47  &  7.11  &  2.92  &  5.90 &  5.90 &  6.53  \\
\hline
 AC  DWT &3.64 & 7.27 & 7.26 &7.90  &  3.31 &  6.61 & 6.64 & 7.29  & 3.01  & 6.06 & 6.05 & 6.70  \\
\hline
 symlet DTT &3.99 & 7.22 &7.25 & 7.63  & 3.64 &  6.65 & 6.66 & 7.05  &3.31 & 6.11 &6.12 & 6.50 \\
\hline
 DT MLT &  4.30  &  8.01 &  7.74 &  8.13  &  3.95  &  7.40 & 7.12 &  7.53 &   3.62 &  6.82 &  6.53 &  6.96 \\
\hline
 AC  DTT  &\textbf{4.39} &\textbf{ 8.04} &\textbf{ 7.83} &\textbf{ 8.24}  & \textbf{4.02} & \textbf{7.44} &\textbf{ 7.20} & \textbf{7.64}  & \textbf{3.68} &\textbf{ 6.85} & \textbf{6.60} &\textbf{7.05}  \\
\hline
\end{tabular}
\end{center}
\caption{Denoising results for different initial SNR's and different wavelets families. The three previous images are studied. The considered estimators are SureShrink (SURE) \cite{Donoho_Johnstone1995}, Neighblock (NB) \cite{Cai_Silverman2001} , Bivariate Shrinkage (Biv) \cite{Sendur_Selesnick2002}  and Visushrink (Visu). \label{tab:compare_ond}}
\end{table}